\documentstyle[aps,psfig]{revtex}
\begin{document}
\title{CMB Anisotropies in the Presence of Extra Dimensions}
\author{C.~S.~Rhodes${}^1$, C.~van~de~Bruck${}^2$, 
Ph.~Brax$^3$ and A.--C.~Davis${}^{1}$}
\address{$^1$Department of Applied Mathematics and Theoretical
Physics, Centre for Mathematical Sciences, University of Cambridge \\
Wilberforce Road , Cambridge CB3 0WA, U.K. \\
$^2$ Astrophysics Department, Oxford University, Keble
Road\\ Oxford OX1 3RH, U.K.\\
$^3$ Service de Physique Theorique, CEA--Saclay \\
F-91191 Gif/Yvette Cedex, France}
\maketitle
\begin{abstract}
We discuss the effect of the time evolution of extra dimensions on CMB
anisotropies and large-scale structure formation.  We study the impact
of scalar fields in a low-energy effective description of a general
class of brane world models on the temperature anisotropy power
spectrum. We show that when the coupling between these scalar fields
and matter evolves over cosmological timescales, current observations
of the CMB anisotropies can constrain primordial values of the fields
in a manner complementary to local, late-time tests of gravity.  We
also present the effect of these fields on the polarization anisotropy
spectra and the growth of large-scale structure, showing that future
CMB observations will constrain theories of the Universe involving
extra dimensions even further.
\end{abstract}

\vspace{0.25cm}

PACS--Nr.: 98.80.-k, 98.80.Es, 04.50.+h, 11.25.Wx

\section{Introduction}

The dominant force in the universe is gravity. According to modern
cosmological theories, primordial density perturbations were generated
in the very early universe by quantum fluctuations, whereas gravity
shaped the structures of the universe
\cite{liddlelyth}--\cite{dodelson}.  This standard picture of
structure formation by gravitational instability has gained remarkable
support from observations of the fluctuations in the cosmic microwave
background (CMB) radiation \cite{WMAP}. In addition, predictions of
hierarchical clustering on scales up to 1000 Mpc in the context of
cold dark matter theories (CDM) are in very good agreement with
observations of the large scale structures in the universe.  These
observations may be used to determine the present cosmological
parameters, such as the density of cold dark matter, the age of the
universe and the equation of state of dark energy among others.

These observations allow us not only to determine cosmological
parameters, but also to test our theories of gravity itself. So far,
no observation of a deviation from General Relativity (GR) has been
reported, but, as we shall see, observations in the CMB and the large
scale structures (LSS) provide important complementary tests to
experiments on Earth or in the Solar System which constrain deviations
from GR.  The growth of perturbations predicted by a given theory of
gravity is sensitive to the details of the theory; therefore, the
study of perturbations can give important insights about any
deviations from GR at different cosmological epochs.

One well-known class of models where deviations from GR are predicted
is that of scalar--tensor theories, in which the gravitational sector
contains not only a tensor field but also scalar fields.  The
Brans--Dicke theory is an example of such a scalar--tensor theory (see
\cite{fujiimaeda} for a recent review on scalar--tensor theories).
One particular feature of these models is that Newton's constant is no
longer constant in the cosmological evolution; alternatively, in the
so--called Einstein frame, which can be obtained from the original
theory after a conformal transformation and field redefinitions,
Newton's constant is truly constant, but the masses of particles are
no longer independent of the space-time coordinates.  Calculations of
CMB anisotropies in scalar--tensor theories
\cite{kamionkowski}--\cite{chiba} show that the positions and
amplitudes of the acoustic peaks usually depend on the parameters of
the theory.  There is no general trend on how the peaks are affected,
because the dynamics of the fluctuations depend critically on the
evolution of the scalar field and its coupling to matter; but the
positions and amplitudes of all peaks are affected.  Similarly, the
slope and amplitude of the matter power spectrum are affected. In
\cite{kamionkowski}, \cite{uzan} and \cite{chiba} certain models of
scalar--tensor theories and their cosmological consequences are
discussed in considerable depth. In \cite{amendola} and \cite{bean}
models are discussed in which the scalar field plays the role of dark
energy and couples to dark matter. In \cite{zahn} it is shown that CMB
anisotropies can probe deviations from the standard Friedmann
equation.

The studies of scalar--tensor theories and their cosmological
consequences are of considerable interest, because theories beyond the
Standard Model of particle physics imply that GR is not fundamental.
Prime examples of theories in which deviations from GR are expected
are those involving supergravity and superstrings (or their extension:
M--theory) \cite{polchinski}.  In particular, models which predict the
existence of extra dimensions can at low energies usually be described
as scalar--tensor theories.  Typically, the low energy effective
description of higher dimensional theories contains many light scalar
fields (so--called moduli fields), which couple to matter in a manner
dependent on the details of the higher dimensional theory.  In this
paper we shall discuss the scalar fields appearing in brane world
models and their effect on the evolution of cosmological perturbations
(for recent reviews on brane worlds see, for instance,
\cite{braxvandebruck}--\cite{maartens} and references therein).  The
brane world model we consider is of the two brane kind. The model is
quite generic and contains the well--known Randall--Sundrum brane
world model (RS I) \cite{RSI} as a special case.  As it turns out (see
section \ref{sec:Effective} below), the  model predicts the
existence of two scalar fields at low energies. One of the scalar
degrees of freedom is associated with a bulk scalar field, i.e.\ a
scalar field which can propagate in the higher dimensional spacetime
between the branes.  The other scalar degree of freedom is related to
the physical distance between the branes.  The two scalar fields
couple to matter {\it
  differently} and thereby affect the growth of perturbations in the
universe each in their own way.

In the theory to be discussed, the coupling function of one of the
scalar fields is constant, its value depending on only one free
parameter of the higher-dimensional theory.  This parameter has to be
chosen to be small in order for the theory to be consistent with local
(Solar System) experiments.  The other coupling function, however,
depends on the second field itself.  Barring any stabilising
mechanism, this field evolves in time due to the cosmological
evolution.  Thus, even though the coupling must be small today, it
could have been larger in the early universe -- in particular at the
epoch of matter--radiation decoupling.  Interestingly, in our theory
the matter coupling of the second field is driven toward small values
during the matter dominated epoch, meaning that the theory can easily
be made consistent with experiments in the solar
system\footnote{Attractor--like behaviour has been found in other
scalar--tensor theories as well \cite{damour}.}.

In this paper we discuss the impact of the moduli fields on the
evolution of cosmological perturbations.  We compare the predictions
to the $\Lambda$CDM model.  We furthermore assume that there is no
potential energy for the scalar fields, so that the energy density of
the scalar field is always much smaller than radiation or matter
energy density. In the models used in this paper, the reported
acceleration of the universe at low redshift is always caused by a
cosmological constant.  This ensures that the distance to the last
scattering surface and the time of matter--radiation equality is only
marginally affected by the presence of the scalar fields.
Furthermore, in order to distinguish between the traces of the
different fields we study them individually.

The paper is organised as follows: In section \ref{sec:Effective} we
briefly review the five dimensional setup of a brane world theory and
discuss the effective theory in four dimensions; readers familiar
with brane worlds or not interested in the details may skip this part.
In section \ref{sec:Cosmo} we write down the action to be discussed in
this paper and derive the field equations.  We also study the
cosmological background evolution of the fields.  In section \ref{sec:Pert} we
derive the perturbation equations and discuss the influence of the
individual fields on CMB anisotropies and the matter power spectrum,
and we conclude in section \ref{sec:Conc}.  This paper is addressed to readers with
different backgrounds, so we have attempted to be comprehensive, and have
included some pedagogical explanations  to highlight  salient points; interested readers should of course
consult   the various cited review articles  for further
clarification and discussion.

\section{A Five Dimensional Theory and its Effective Field Theory 
at Low Energies}
\label{sec:Effective}

As an introduction to the theory studied in this paper, we begin with
a discussion of a higher--dimensional theory motivated from brane
worlds. In these kind of theories, the standard model particles (and
maybe also some other, exotic, form of matter) are confined on a
four\footnote{Three space and one time.}--dimensional object (called a
brane), which moves in a higher--dimensional space (known as the
bulk).  The brane itself carries some intrinsic tension as well.  In
our model, the bulk spacetime is five--dimensional.  This setup is
well motivated by recent developments in string theory and provides a
useful starting point for more complicated models \cite{lukas}.  Also
extracted from string theory is the fact that there is a second brane
somewhere in the bulk.  Confined on this second brane, which has an
intrinsic tension too, can be some form of matter whose only direct
interactions with the matter on first brane are mediated via gravity.
This matter is a potential dark matter candidate.

\begin{figure}[!ht]
\hspace{3.5cm}\psfig{file=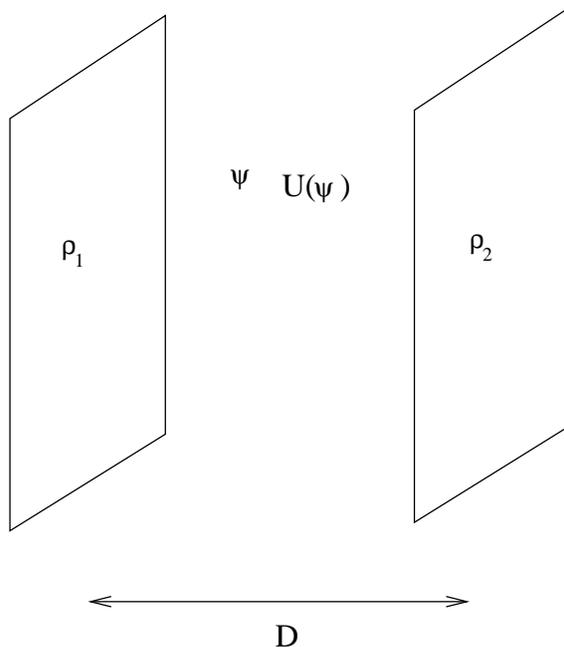,width=7.5cm}
\caption[h]{
  The setup of the two brane model. $\psi$ denotes the bulk scalar field 
  (with potential energy $U(\psi)$),
  which can propagate between the branes. $\rho_1$ and $\rho_2$ denote
  the total (i.e. tension + matter) energy density confined on the
  individual branes. $D$ is the physical distance between the branes.}
  \label{fig:setup}
\end{figure}

The physical spacetime stretches between the two branes, so that the
branes form the boundary of the five--dimensional spacetime.  The
space between is not empty; in general, some form of matter is
expected to propagate through the bulk: in the model under
consideration, a scalar field is present.  Taking this scalar field
into account, the overall model setup is presented in figure
\ref{fig:setup}.

For energy densities much higher than the brane tension it has been
noted \cite{binetruy} that the expansion of the four dimensional brane
universe is not governed by the usual four--dimensional Einstein
equations, according to which the expansion rate $H$ is proportional
to the matter density $\rho$; instead, it was found that the
dependence of $H$ on $\rho$ is $H \propto \rho$.  This behaviour was
observed in one--brane scenarios \cite{binetruy} but was shown to hold
in two--brane scenarios as well (see in particular
\cite{boundaryinflation} and \cite{clinevinet}).  However, for
densities much less than the brane tension, the usual law
$H\propto\sqrt{\rho}$ can be recovered, albeit with some corrections
coming from the scalar fields associated with the bulk dynamics.

As already mentioned in the introduction, the effective field theory
at low energies (i.e.\ at energies much less than the brane tension)
can be written as a scalar--tensor theory with two scalar fields.  The
action of the theory was derived in \cite{modulipaper} and reads
\begin{eqnarray}\label{fundamental}
S_{\rm EF} &=& \frac{1}{16\pi G_N} 
\int d^4x \sqrt{-g}\left[ {\cal R} -  \frac{12\tilde \alpha^2}{1+2\tilde\alpha^2}
(\partial \varphi)^2 - \frac{6}{2\tilde \alpha^2 + 1}(\partial R)^2 
- V(\varphi,R)\right].
\end{eqnarray}

In this equation, ${\cal R}$ is the Ricci scalar, $\varphi$ and $R$
are the two scalar fields and $V$ is the total potential energy of the
fields.  $\tilde \alpha$ is a parameter of the higher-dimensional theory,
which will be left free, though note that models of particle physics,
in general, will make predictions about its value \cite{braxdavis}.
In the Einstein frame, each individual matter form on the branes
usually couples to the fields $\varphi$ and $R$, so that the action
has the general form
\begin{equation}
S_{\rm Matter} = 
S_{{\rm Matter},1}(\psi_1, A(\varphi,R)g_{\mu\nu}) + 
S_{{\rm Matter},2}(\psi_2, B(\varphi,R)g_{\mu\nu}),
\end{equation}
where $\psi_1$ denotes the matter form on brane 1 and $\psi_2$ 
the matter form on brane 2. The functions $A$ and $B$ in general 
depend on both fields and are not equal. 

The interpretation of $\varphi$ and $R$ is not straightforward, but
the fact that there must be two scalar degrees of freedom at low
energies can be understood as follows: first, the bulk scalar field
can propagate along the brane directions, because it depends on all
spacetime coordinates.  The other degree of freedom which appears in
the low energy effective action is related to the distance between the
branes, which is a scalar function of the four spacetime coordinates.
We note furthermore that the case where $\tilde\alpha$ is zero
(corresponding to the second, field-dependent coupling we consider
below) subsumes the Randall--Sundrum two brane model, where there is
no bulk scalar field.

The matter couplings are determined by the functions $A$ and $B$, or,
to be more precise, by their derivatives with respect to $\varphi$ and
$R$. For the brane world model we consider, the coupling functions are
given by (see \cite{modulipaper} for details)
\begin{eqnarray} 
\beta_\varphi^{(1)} &\equiv& 
\frac{\partial \ln A}{\partial \varphi} =  
-\frac{2\tilde\alpha ^2}{1+2\tilde\alpha ^2}, \hspace{0.5cm}
\beta_\varphi^{(2)} \equiv \frac{\partial \ln B}{\partial \varphi} =
-\frac{2\tilde\alpha ^2}{1+2\tilde\alpha ^2}, \label{coupling1} \\
\beta_R^{(1)} &\equiv& 
\frac{\partial \ln A}{\partial R} = 
\frac{\tanh R}{1+2\tilde\alpha ^2}, \hspace{0.5cm}
\beta_R^{(2)} \equiv \frac{\partial \ln B}{\partial R} = 
 \frac{(\tanh R)^{-1}}{1+2\tilde\alpha ^2}. \label{coupling2}
\end{eqnarray}
The functions $\beta$ depend on the details of the higher dimensional
theory, as they provide all information about the nature of the
higher--dimensional space (such as its geometry or curvature) relevant
at low--energy.  Throughout this paper we shall consider coupling
functions of the form above.

There are different ways of obtaining the effective theory at
low energies. In \cite{modulipaper}, the moduli space approximation
was used, in which one integrates the heavy Kaluza--Klein like
excitations out. Strictly speaking, the action above is only valid if
the Kaluza--Klein excitations, corresponding to massive particles in
four dimensions, are negligible. For example, massive partners of the
gravitational zero mode, described by $g_{\mu\nu}$, are neglected.
Another method to derive the low--energy effective action was used in
\cite{kannosoda}--\cite{shiromizu}, which is based on solving the
five--dimensional Einstein equation to linear order, and then
projecting onto the brane on which the standard model particles are
confined. The theory is then written in the Jordan frame.  This method
is widely used in works on brane cosmology and has the advantage in
that it provides a geometrical framework for  the full five--dimensional
problem\footnote{In Ref. \cite{koyama} this formalism was used to calculate 
the CMB anisotropies in the one--brane scenario of Randall and Sundrum 
\cite{RSII}.}. However, it should be noted that the action with the two
methods lead to the same low--energy effective theory: after a
transformation to the Einstein frame, the action obtained by the
projective approach is equivalent to the action (\ref{fundamental}).

In practical terms, the use of the moduli space approximation limits
the backwards extrapolation in time; for the use of the moduli space
approximation to be valid, we require that the Hubble scale be much
greater than the scale under consideration.  So, $H(z) \approx (1+z)^2
H_0$ must be much less than $k^{-1}$, the size of the extra dimension,
so we can see for instance that millimetre (about $10^{13}{\rm
GeV}^{-1}$) scales can be dealt with using this approximation to
redshifts of about $10^{15}$ (while numerical integration, of which
more discussion in section \ref{sec:Pert}, typically starts at
redshifts of $10^9$). As already said, in deriving the action above,
one assumes that Kaluza--Klein excitations are negligible at these
redshifts. For example, the bulk scalar field in this approximation
does not depend on the extra dimension. We further assume that such
excitations are not produced at the epochs which we are interested
in. This should be valid because any excitation corresponds to a heavy
particle.

Clearly this theory, described by the action in equation
(\ref{fundamental}), is different from GR, and therefore gravitational
experiments constrain parameters of the theory. In order for the
theory to agree with experiments on earth and observations in the
solar system, the parameter $\tilde\alpha $ has to be rather small
($\tilde\alpha \le 10^{-2}$) and the field $R$ today, should be small
($R\le 10^{-2}$), too.  However, whereas the parameter $\tilde\alpha $
has to be chosen small from the beginning, the field $R$ is dynamical
and so one has to choose its initial condition such that it is small
today.  Given this, there is another, more intriguing possibility:
that during the cosmological history the field is driven towards small
values.  The cosmological evolution of the system was discussed in
\cite{modulipaper}, where it was shown that $R$ is indeed driven
toward small values in a stable fashion.  Apart from local
experiments, the theory is constrained by cosmological considerations.
Firstly, nucleosynthesis bounds have to be respected, which put
constraints on both $\tilde\alpha $ and the initial value of the field
$R$. It is beyond the scope of the paper to constrain the theory by
nucleosynthesis as a detailed analysis with two fields involved has to
be performed. The second important cosmological constraint on the
fields is enforced by CMB anisotropies and large scale structure
considerations. We will in this paper develop an understanding of the
how the CMB is influenced by the fields.

Before we discuss perturbations in models of the form above, we
comment on the frame to be used.  The issue of the frame and its
physical relevance is not new, and has been hotly discussed in the
literature (see e.g. \cite{conformal}).  The reason for choosing the
Einstein frame instead of the Jordan frame is that, written in the
Jordan frame, one the of the scalar fields has a negative kinetic
energy \cite{modulipaper} (the metric in the Jordan frame is the
induced metric of the first brane, where the standard model particles
live).  However, this is not a signal of an instability, because in
the Einstein frame the kinetic term is positive. The fact that the
energy density of the scalar field is not positive definite in the
Jordan frame is not unique to field theories derived from brane worlds
but appears often in scalar--tensor theories.  Several authors have
used this as an argument for using the Einstein frame as the physical
frame (which is unique), and we follow this argument here. We would
like to point out, however, that in the case of the field with
non--constant coupling, the matter coupling of this field is very
small today, so that at the present epoch the theory is
indistinguishable from General Relativity.

\section{Cosmology and Moduli Fields}
\label{sec:Cosmo}

As  discussed in the last section, there are usually several
scalar fields in the low--energy effective action, each of which
couples to matter.  The goal of this paper is to understand the
effects of the {\it individual fields} on the evolution of
perturbations.  We shall henceforth discuss them separately, dealing
with only one scalar field at a time.  We shall see in the next
section that there are differences in the predicted CMB power spectrum
due to the fact that the field coupling function is in one case
constant whereas in the other case depends on the field.  We shall
study two cases individually: the matter coupling will either be of
the form (\ref{coupling1}) or ({\ref{coupling2}}).

The toy model we consider in this paper is described by the following 
action:
\begin{eqnarray}\label{action}
S = \int d^4x \sqrt{-g} \left[ \frac{{\cal R}}{2\kappa^2} 
-g^{\mu\nu}(\partial_\mu \phi)(\partial_\nu \phi) - V(\phi)\right]
+ S_V(\psi_V, A(\phi) g_{\mu\nu}) 
+ S_{IV}(\psi_{IV}, B(\phi)g_{\mu\nu})
\end{eqnarray}
In the context of brane worlds, the index $V$ stands here for matter
on brane 1, whereas $IV$ stands for matter on brane 2. The functions
$\psi_V$ and $\psi_{IV}$ stand for the matter fields on each
individual branes. The field $\phi$ in equation (\ref{action}) takes the
role of either $\varphi$ or $R$ in the last section. In the following 
we shall work with reduced Planck units $\kappa\equiv 1$.

From the action above, one can derive the following field equations:
\begin{equation} \label{Einstein}
G_{\mu\nu} = \left(T_{\mu\nu}^{(V)} 
+ T_{\mu\nu}^{(IV)} + \phi_{,\mu}\phi_{,\nu} - 
\frac{1}{2}g_{\mu\nu} (\partial \phi)^2 + V(\phi)\right)
\end{equation}
\begin{equation} \label{Kleingordon}
\Box\phi = -\beta_{V} T^{V} - \beta_{IV} T^{IV}  
\end{equation}
\begin{equation}\label{Energy}
T_{\mu}^{(i)\nu;\mu} = \beta_{(i)} T^{(i)} \phi^{,\nu}, \hspace{1cm} 
{\rm with }\hspace{0.2cm}{\it i} = ({\it V,IV}).
\end{equation}
In these equations, $T$ stands for the trace of the energy--momentum 
tensor for each individual matter. The functions $\beta$ are 
given by 
\begin{equation}
\beta_1 = \frac{\partial \ln A}{\partial \phi}, \hspace{0.5cm}
\beta_2 = \frac{\partial \ln B}{\partial \phi}.
\end{equation}

In the first case under consideration, the functions $\beta$ do not
depend on the scalar field (as it is the case with one of the scalar
field in section \ref{sec:Effective}) and choose $\beta$ to be the form
\footnote{Note that $\alpha$ in this expression is different from 
$\tilde \alpha$ in Section 2, because of the field redefinition.}
\begin{equation}
{\rm Case}\hspace{0.2cm}{\rm I}: \hspace{0.2cm} A = B \propto \exp(-2\alpha^2\phi/(1+2\alpha^2)) 
\Rightarrow \hspace{0.5cm} 
\beta_1 = -\frac{2\alpha^2}{1+2\alpha^2} = \beta_2.
\end{equation}
We have retained the main features of the brane model, i.e. $\beta_{1,2}$ is negative and bounded.
In the second case (corresponding to the field $R$ in the last 
section), the functions $\beta$ depend on the field and we have
\begin{equation}
{\rm Case}\hspace{0.2cm}{\rm II}: \hspace{0.2cm} A \propto \cosh(\phi), \hspace{0.2cm} 
B \propto \sinh(\phi)
\Rightarrow \hspace{0.5cm} 
\beta_1 = \tanh(\phi) = 1/\beta_2.
\end{equation}
As before we have kept the same functional dependence as in the brane models. 

For a homogeneous and isotropic universe with flat spatial sections, 
the line element reads
\begin{equation}
ds^2 = a^2(\tau)\left(-d\tau^2 + \delta_{ij} dx^i dx^j\right).
\end{equation}
The field equations read
\begin{equation}
H^2 = \frac{1}{3}a^2 \left(\rho_V + \rho_{IV} + \frac{1}{2a^2}\dot\phi^2 
+ V(\phi)\right),
\end{equation}
\begin{equation}\label{Kleingordon2}
\ddot \phi + 2H\dot\phi + a^2 \frac{\partial V}{\partial \phi} 
= - \beta_{V}(\rho_V - 3p_V)a^2 - \beta_{IV}(\rho_{IV} - 3p_{IV})a^2
\end{equation}
\begin{equation}\label{enercon}
\dot \rho_{(i)} + 3H(\rho_{(i)} + p_{(i)}) = 
\beta_{(i)} \left(\rho_{(i)} - 3 p_{(i)}\right)\dot\phi .
\end{equation}
In these equations, the dot represents a derivative with respect to
the conformal time $\tau$ and $H=\dot a /a$. Note that in
equations (\ref{Kleingordon}) and (\ref{Kleingordon2}) we must sum over
all matter forms. In this paper we will assume that $T^{IV}_{\mu\nu} = 0$. 
This would correspond to no matter on the second brane in the theory 
of the last section. 

The theory above differs from General Relativity and is therefore
constrained by observations. Firstly, local experiments (i.e. on earth
and in the Solar System) constrain the post--Newtonian parameters
$\gamma_{\rm PN}$ and $\beta_{\rm PN}$ given by
\begin{eqnarray} 
\gamma_{\rm PN} - 1 &=& -2\frac{\beta_{\rm A/B}^2}{1 + \beta_{\rm A/B}^2}, \\
\beta_{\rm PN} - 1 &=& \frac{1}{2} \frac{\beta_{\rm A/B}^2}{(1 + \beta_{\rm A/B}^2)^2}
\frac{d\beta_{\rm A/B}}{d\phi},
\end{eqnarray}
where $\beta_{\rm A/B}$ stand for the functions $\beta_1$ for the
cases I or II, respectively.  Current constraints give \cite{Will}
\begin{equation}\label{PPN}
|\gamma_{\rm PN} - 1| \leq 2\cdot 10^{-3},\hspace{0.5cm} 
|\beta_{\rm PN} - 1| \leq 6 \cdot 10^{-4}.
\end{equation}
Nucleosynthesis constraints limit the effective number of degrees 
of freedom for relativistic particles at this epoch. In general 
the constraints have to be worked out in detail, but there is a 
simple way to get a rough limit on the theory from nucleosynthesis 
\cite{bartolo}. Because 
the energy conservation equation (\ref{enercon}) implies that 
$\rho_{\rm M} a^3 \neq {\rm constant}$ in general, the 
expansion rate at nucleosynthesis is different from its value 
in General Relativity. This leads to \cite{bartolo} 
\begin{equation}\label{nucleosynthesis}
\left(\frac{A(\phi_{\rm nuc})}{A(\phi_0)}\right)^2 \leq 1.2.
\end{equation} 

Although we have derived the general equations, from now on we shall
set $V(\phi) = \Lambda$, because we wish to compare the results for
the perturbations with the $\Lambda$CDM model.  Before we turn our
attention to cosmological perturbations and their evolution, however,
we briefly study the evolution of the background in the matter
dominated era for the case A and B separately.

\begin{center}
\underline{Case I: The case for constant coupling parameter $\beta$}
\end{center}

The model in this case has similarities to the model discussed in
\cite{amendola2}, although we do not have an exponential potential for
the field, and so the critical points found in \cite{amendola2} do not
apply here.  In the radiation dominated epoch, the field is almost
constant, because the terms on the left--hand side on the
Klein--Gordon equation (\ref{Kleingordon2}) dominate.  The constraints (\ref{PPN}) on the post--Newtonian parameter lead to 
$\alpha \leq 0.1$. 
We can easily find the solutions to the background
equations in the matter dominated epoch:
\begin{eqnarray}
a(\tau) &=& \left(\frac{\tau}{\tau_0}\right)^x, \hspace{0.1cm}{\rm with} \hspace{0.5cm} 
x = 2 - 8\alpha^4 \nonumber \\
\phi &=& \phi_0  + 4\alpha^2 \ln\left(\frac{\tau}{\tau_0}\right) \label{phimatter}.
\end{eqnarray}
Therefore, the universe expands a little slower than expected,
although the correction is very small.  An important point has to be
made here, namely that matter does not scale like $\rho_{\rm matter}
\propto a^{-3}$, but rather like
\begin{equation}\label{matterscaling}
\rho_{\rm matter} \propto a^{-3} \exp\left(\int \beta d\phi \right).
\end{equation}

For the theory under consideration here, this implies that, for a
given matter density today, the matter density in the past was greater
than it would have been under standard cosmology.

\begin{center}
\underline{Case II: The case for non-constant coupling parameter $\beta$}
\end{center}
The low--energy effective theory now corresponds to the first model by
Randall and Sundrum \cite{RSI}. As we will see, there is an attractor
mechanism at work, which drives the field towards small values and
thereby generating small couplings between $\phi$ and matter, as it
is dictated by observations on Earth and in the Solar System.

The equations are difficult to handle analytically.  Assuming that $\phi$
is small and slowly varying at the onset of matter domination, one can
find that
\begin{equation}
a(\tau) = \left(\frac{\tau}{\tau_0}\right)^2, 
\end{equation}
and that $\rho_{\rm matter} \propto a^{-3}$. The 
field $\phi(\tau)$ now goes like
\begin{equation}
\phi(\tau) = A \frac{\cos\left(\sqrt{\frac{39}{4}}\ln \left(\tau/\tau_0\right)\right)}{(\tau/\tau_0)^{3/2}}
+ B \frac{\sin\left(\sqrt{\frac{39}{4}}\ln \left(\tau/\tau_0\right) \right)}{(\tau/\tau_0)^{3/2}}.
\end{equation}
Thus, apart from oscillating behaviour, the field $\phi$ in this case
decays during matter domination.

Although the solutions above are based on a crude approximation, this
qualitative behaviour is found even if the field value was initially
larger \cite{modulipaper}.  For large values of $\phi$, however,
equation (\ref{matterscaling}) is a better approximation.

Thus, although it is easy to fulfill the constraints (\ref{PPN}) due to 
the cosmological attractor, the nucleosynthesis constraints give a limit 
on $\phi(\tau=\tau_{\rm nucl})\leq 0.4 $. 

\section{Cosmological Perturbations and Moduli Fields}
\label{sec:Pert}

In the following we discuss the evolution of perturbations in the
presence of moduli fields.  This section is organized as follows:
first we will write down the perturbation equations. Then we will
solve them (approximately) in the matter dominated era, in order to
gain some understanding about the effects of the coupling between the
field and matter.  Finally we discuss the solutions of the numerical
computation of the spectrum of anisotropies in the CMB and the matter
power spectrum.

\subsection{The Perturbation Equations in the Synchronous Gauge}
\label{sec:PertEqs}

We work in the synchronous gauge (using the notations as in
\cite{mabertschinger}); i.e.\ the perturbed line element has the form
\begin{equation}
ds^2 = a^2(\tau)\left(-d\tau^2 + (\delta_{ij}+h_{ij}) dx^i dx^j\right).
\end{equation}
After Fourier transformation, the perturbed Einstein equations read:
\begin{eqnarray}
2k^2 \eta - \frac{\dot a}{a} \dot h 
&=&  a^2 \delta T^{0}_{~0} \label{einsteinone}\\ 
\ddot h + 2\frac{\dot a}{a}\dot h - 2k^2\eta 
&=& - a^2 \delta T^{i}_{~i} \label{einsteintwo},
\end{eqnarray}
where $h$ and $\eta$ are defined by ($\hat{\bf k}$ is the unit vector in 
direction of $\bf k$)
\begin{equation}
h_{ij} = \int d^3k e^{i{\bf k} \cdot {\bf x}} \left[
{\bf \hat k}_{i}\cdot{\bf \hat k}_{j} h({\bf k},\tau) + 
\left( {\bf \hat k}_{i}\cdot {\bf \hat k}_{j} 
- \frac{1}{3}\delta_{ij}\right) 6\eta({\bf k}, \tau) \right]
\end{equation}

>From the energy--momentum conservation equation one obtains a set of
two equations for the evolution of the density contrast and the
divergence of the velocity field.  Defining
\begin{eqnarray}
(\rho+p) \theta &=& ik^j \delta T^{0}_{~j} \\
(\rho+p) \sigma &=& -\left(\hat k_i \cdot \hat k_j 
- \frac{1}{3}\delta_{ij}\right)\left( T^{ij} - \delta^{ij} T^k_{~k} /3 \right)
\end{eqnarray}
and writing $p_{i}= w_{(i)} \rho_{(i)}$, 
$c_{s(i)}^2 = \partial p_{(i)} / \partial \rho_{(i)}$ 
and $\theta = ik^i v_i$ we find from the 
evolution equation for the density contrast $\delta_{(i)} 
= \delta \rho_{(i)}/\rho_{(i)}$ 
\begin{eqnarray}
\dot \delta_{(i)} &=& - \left(1+w_{(i)}\right)\left[ \theta_{(i)} 
+ \frac{\dot h}{2} \right] - 3\frac{\dot a}{a} \left( c_{s(i)}^2 -
w_{(i)}\right)\delta_{(i)} \nonumber \\
&+&\frac{\partial \beta_i}{\partial\phi}\left(1-3w_{(i)}\right)\dot \phi \delta
\phi + \beta_i \left(1-3w_{(i)}\right) \dot{(\delta \phi)}
- 3\dot\phi \beta_i \left(c_{s(i)}^2 - w_{(i)}\right)\delta_{(i)}. \label{momentum1}
\end{eqnarray}
The divergence of the velocity field $\theta_{(i)} = v^{(i)j}_{~~~~,j}$ of
each individual fluid evolves according to
\begin{eqnarray} 
\dot\theta_{(i)} &=& -\frac{\dot a}{a}\left( 1-3w_{(i)} \right)
\theta_{(i)} - \frac{\dot w_{(i)}}{1+w_{(i)}}\theta_{(i)} 
+ \frac{c^2_{s(i)}}{1+w_{(i)}} k^2 \delta - k^2\sigma_{(i)} \nonumber \\
&+& \beta_i\frac{1-3w_{(i)}}{1+w_{(i)}} k^2 \delta\phi 
- \beta_i (1-3w_{(i)}) \dot\phi \theta_{(i)}. \label{momentum2}
\end{eqnarray}

It is important to note that these equations reduce to those of
ordinary General Relativity, if the fluid under consideration has an
radiation--like equation of state, i.e.\ if $w_{(i)} = 1/3$. This
means that the equations governing the dynamics of photons and
relativistic neutrinos are those found in General Relativity.
Important changes occur in the equations for CDM and baryons:

\begin{center}
{\it \underline{Cold Dark Matter:}}
\end{center}
In this case, we have $w=0=c_s$ and vanishing anisotropic stress. Thus
\begin{equation} \label{cdmone}
\dot\delta_{c} = - \left[ \theta_{c} + \frac{\dot h}{2} \right] 
+ \frac{\partial \beta_{c}}{\partial\phi}\dot \phi \delta\phi 
+ \beta_{c} \dot{(\delta \phi)}
\end{equation}
\begin{equation}\label{cdmtwo}
\dot\theta_{c} = -\frac{\dot a}{a}\theta_{c}
+ \beta_{c} k^2 \delta\phi - \beta_{c} \dot\phi \theta_{c}
\end{equation}

\begin{center}
{\it \underline{Baryons:}}
\end{center}
In this case, we have $w=0=c_s$ and vanishing anisotropic stress (a
very good approximation after neutrino decoupling
\cite{mabertschinger}).  As in \cite{mabertschinger}, we do not
neglect the quantity $c_s^2 k^2 \delta$, which becomes important on
small length scales and thus large $k$. We also have to add the terms
describing the momentum transfer between photons and baryons due to
Thomson scattering.  Therefore, we have
\begin{equation}
\dot\delta_{b} = - \left[ \theta_{b} + \frac{\dot h}{2} \right] 
+ \frac{\partial \beta_b}{\partial\phi}\dot \phi \delta\phi 
+ \beta_b \dot{(\delta \phi)}
\end{equation}
\begin{equation}
\dot\theta_{b} = -\frac{\dot a}{a}\theta_{b} + c_s^2 k^2 \delta_b
+ \frac{4\rho_\gamma}{3\rho_b}an_e\sigma_T(\theta_\gamma - \theta_b)
+ \beta_b k^2 \delta\phi - \beta_b \dot\phi \theta_{b}
\end{equation}
where $n_e$ is the number density of free electrons and $\sigma_T$ the 
Thomson cross section. 

The only new terms compared to General Relativity are the ones
proportional to $\beta$ and its derivative. Note that we have to add
an evolution equation for $\theta_{c}$: while we can still set
$\theta_{c}=0$ initially (in order to fix the synchronous gauge
completely), in General Relativity, where there is no extra field, it
stays zero; however, fluctuations in $\phi$ are a source and
therefore, in the theory we consider, it evolves according to the
above equation.

Finally, the equation for the fluctuations in $\phi$ evolve 
according to 
\begin{equation}\label{kgpertur}
\ddot{(\delta \phi)} + 2\frac{\dot a}{a}\dot{(\delta \phi)} + 
\left(k^2 + a^2 \frac{\partial^2 V}{\partial \phi^2} \right)\delta \phi
+ \frac{1}{2}\dot h\dot\phi =-\left(\beta_{c} \rho_{c}\delta_{c} 
+ \beta_{b} \rho_{b}\delta_{b} 
+ \frac{\partial \beta_b}{\partial \phi}\rho_b(\delta \phi) 
+ \frac{\partial \beta_{c}}{\partial \phi}\rho_{c}(\delta \phi)\right)a^2
\end{equation}
These equations are quite general. Again, we are assuming 
$V=\Lambda = {\rm const.}$

\subsection{Solutions to the Perturbation Equations in the Matter Dominated Era}
Before we give the results for the CMB anisotropies and matter power
spectrum, we try to gain some analytical insight from the perturbation
equations in the matter dominated epoch on sub-horizon scales. We
neglect radiation and the baryons, taking only into account CDM and
the scalar field in this regime.  Furthermore, we consider only the
case for the field with constant coupling, which is tractable
analytically.

Taking the derivative of equation (\ref{cdmone}) and making use of 
equations (\ref{einsteinone}), (\ref{einsteintwo}), 
(\ref{cdmtwo}) and the Friedmann equation, one gets 
\begin{equation}
\ddot \delta_c + H\dot \delta_c -\frac{3}{2}H^2 \delta_c 
= 2\alpha^2 k^2 \delta \phi + 2\dot\phi\dot{(\delta\phi)}
- 2\alpha^2 \left(H \dot{(\delta \phi)} + \ddot{(\delta \phi)}\right) 
- 2\alpha^2 \dot\phi \theta_c.
\end{equation}
Using the perturbed Klein--Gordon equation, ignoring oscillations in 
$\delta\phi$ and inserting (\ref{cdmone}) one obtains
\begin{equation}
\ddot \delta_c + (H-2\alpha^2\dot\phi)\dot \delta_c 
- \frac{3}{2}H^2 \left( 1 + 8\alpha^4\right)\delta_c = 0.
\end{equation}
Assuming $a\propto \tau^x$, using the background solution equation 
(\ref{phimatter}) and making the Ansatz $\delta_c \propto \tau^m$, 
one gets up to 
${\cal O}(\alpha^4)$
\begin{equation}
m = 2 + \frac{32}{5}\alpha^4
\end{equation}
for the growing mode. Therefore, compared to the $\Lambda$CDM model, due to the
non--vanishing coupling perturbations in CDM grow {\it faster} than
the scale factor and the transfer function is on small scales bent
towards larger values. We have confirmed this numerically.

\subsection{CMB Anisotropies and the Matter Power Spectrum}
In the following we will discuss the spectrum of anisotropies in the
CMB and the matter power spectrum. It is customary to expand the
temperature anisotropy measured in a given direction ${\bf n}$ in
spherical harmonics according to
\begin{equation}
\frac{\Delta T}{T} ({\bf n}) = \sum_{lm} a_{lm} Y_{lm} ({\bf n}).
\end{equation}
We assume that the density fluctuations are Gaussian random variables, 
as predicted by the inflationary scenario. The statistics of the 
fluctuations are completely specified by the power spectrum $C_l$, defined as 
\begin{equation}\label{Cldef}
\langle a_{lm} a^{*}_{l'm'} \rangle \equiv C_{l} \delta_{ll'}\delta_{mm'}.
\end{equation} 
Let us briefly summarize how the $C_l$s depend on the perturbation
variables and how they are calculated (see \cite{durrer},
\cite{hudodelson} and \cite{dodelson} for excellent detailed reviews on the
physics of CMB anisotropies, \cite{CMBshortbegin}--\cite{CMBshortend} 
are shorter reviews).  The reader familiar with CMB
anisotropies can skip this part.  To illustrate the essential steps
involved in calculating the anisotropy power spectrum, we neglect the
polarization states of the photons.  However, when we present the
results of the calculations, the polarization states have been taken
into account.

At very early times the universe was radiation dominated and the
temperature of the universe was so high, that photons and baryons
formed a plasma. At these high temperatures, photons and baryons are
tightly coupled. On scales less than the horizon the fluid undergoes
acoustic oscillations due to the interplay of gravity, which tries to
compress overdense regions, and the photon pressure, which, when high
enough, acts against the force of gravity.  When the temperature
dropped below 4000 K, electrons and nuclei were able to combine to
atoms (mainly hydrogen and helium) and the mean free path of the
photons became larger than the Hubble horizon. Around that time, the
photons completely decoupled from matter.  On their path, photons may 
experience energy loss due to time--changing gravitational fields (the
so--called integrated Sachs--Wolfe effect, or ISW).  Additionally,
photons may interact with reionized gas (reionization might happen due
to early star formation).

In order to follow the perturbations in the photons correctly, one has
to go beyond the fluid description and study the evolution of the
photon distribution function $f_\gamma$ in phase space, i.e. one has
to solve the Boltzmann equation. In general, $f_\gamma$ is a function
of the comoving position ${\bf x}$, the photon momentum ${\bf p}$ and
conformal time $\tau$. To first order we write $f_\gamma = f_0 +
\delta f_\gamma$, where $f_0$ is the unperturbed photon distribution.
The brightness function $\Delta_T({\bf x},{\bf n},\tau)$ is defined as 
\begin{equation}
\delta f_\gamma ({\bf x},{\bf p},\tau) 
= \left(\frac{\bar T}{4}\frac{\partial f_0}
{\partial \bar T}\right)\Delta_T({\bf x},{\bf n},\tau),
\end{equation}
where $\bar T$ is the unperturbed temperature of the photon gas (i.e.
of the background) and ${\bf n}$ is a unit vector in the direction of
the photon momentum ${\bf p}$.  One can now formulate a perturbed
version of the Boltzmann equation, describing the evolution of
$\Delta_T$. However, it is more useful to perform a Fourier
transform of $\Delta_T({\bf x},{\bf n},\tau)$ and expand the
Fourier coefficients $\Delta_T({\bf k},{\bf n},\tau)$ in Legendre
polynomials (${\bf \hat k}$ is a unit vector in direction of ${\bf
k}$):
\begin{equation}
\Delta_T({\bf k},{\bf n},\tau) = \sum_{l=0}^{\infty} (-i)^l (2l+1) \Delta_{Tl}(k,\tau) 
P_l({\bf \hat k} \cdot \bf n).
\end{equation}
It can be shown that the density contrast in the photon energy
density, $\delta_\gamma$, the divergence of the photon velocity field,
$\theta_\gamma$ and the photon anisotropic stress, $\sigma_\gamma$,
are related to $\Delta_{Tl}$ by
\begin{equation}\label{brightness}
\delta_\gamma = \Delta_{T0},\hspace{0.5cm} \theta_\gamma 
= \frac{3}{4}k\Delta_{T1},\hspace{0.5cm} \sigma_\gamma = \frac{1}{2}\Delta_{T2}. 
\end{equation}
In terms of the moments $\Delta_{Tl}$, the Boltzmann equation becomes
a hierarchy of equations. Using equation (\ref{brightness}), the evolution
equations are given by
\begin{eqnarray}
\dot\delta_\gamma &=& -\frac{4}{3}\theta_\gamma - \frac{2}{3}\dot h, \label{first}\\
\dot\theta_\gamma &=& k^2 \left(\frac{1}{4}\delta_\gamma - \sigma_\gamma \right) 
a n_e \sigma_T \left(\theta_b - \theta_\gamma \right), \\
2\dot\sigma_\gamma &=& \frac{8}{15}\theta_\gamma - \frac{3}{5} k \Delta_{T3} 
+ \frac{4}{15}\dot h + \frac{8}{5}\dot\eta - \frac{9}{5} an_e\sigma_T\sigma_\gamma \label{change} \\ 
\dot \Delta_{Tl} &=& \frac{k}{2l+1}\left[l\Delta_{T(l-1)} - (l+1)\Delta_{T(l+1)} \right] 
- a n_e \sigma_T \Delta_{Tl}, \hspace{1cm}(l\ge 3) \label{last}
\end{eqnarray}
Furthermore, it can be shown that the quantity $C_l$, defined in
equation (\ref{Cldef}), is related to the brightness function $\Delta_{Tl}$
by
\begin{equation}\label{Cl}
C_l = 4\pi \int \frac{dk}{k} \left| \Delta_{Tl} (k,\tau_0) \right|^2 .
\end{equation}

It is straightforward to include polarization. Due to Thomson
scattering, photons are linearly polarized in the plane perpendicular
to ${\bf n}$.  The collision terms in the Boltzmann equation depend on
the polarization state. In order to follow the evolution in phase
space, one defines a {\it total} distribution function $f_\gamma$,
which includes all polarization states and another distribution
function, $g_\gamma$, which is defined as the {\it difference} of the
distribution functions for the individual polarized photons. Defining
the temperature brightness function $\Delta_T$ and the polarization
brightness function $\Delta_P$ similarly to the above procedure, one
can show that $\Delta_T$ obeys the equation
(\ref{first})--(\ref{last}), but equation (\ref{change}) has to be
changed to
\begin{equation}
2\dot\sigma_\gamma = \frac{8}{15}\theta_\gamma - \frac{3}{5} k \Delta_{T3} 
+ \frac{4}{15}\dot h + \frac{8}{5}\dot\eta - \frac{9}{5} an_e\sigma_T\sigma_\gamma
+\frac{1}{10} a n_e \sigma_{T}\left(\Delta_{P0} - \Delta_{P2}\right),
\end{equation}
because the collision term in the Boltzmann equation depends on the
polarization state.  Additionally, one obtains a hierarchy for the
functions $\Delta_{Pl}$:
\begin{equation}
\dot\Delta_{Pl} = \frac{k}{2l+1} \left[l\Delta_{P(l-1)} - (l+1)\Delta_{P(l+1)} \right] 
- a n_e \sigma_T \left[ \Delta_{Pl} + \frac{1}{2}\left( 
\Delta_{T2} + \Delta_{P0} + \Delta_{P2}\right)\left( 
\delta_{0l} + \frac{\delta_{2l}}{5}\right)\right],
\end{equation}
where $\delta_{il}$ in the last term stands for the Kronecker delta. 

The CMB anisotropy power spectrum $C_l$ is still given by equation
(\ref{Cl}).  Thus, in order to calculate the CMB power spectrum $C_l$
(as well as the polarization power spectrum and the
temperature/polarization cross correlation spectrum), one has to
solve a large set of differential equations (the equations for
baryons, CDM and the scalar field presented in section \ref{sec:PertEqs} and the
hierarchy above).  In \cite{husugiyama} solutions to the Boltzmann
equation are found. Here we use an existing code \cite{lewis} and make
the appropriate changes to the evolution of background and
perturbation variables to incorporate the above theory.

\subsubsection{Constant Coupling}
In order to calculate the CMB power spectrum, we make changes to an
existing code \cite{lewis}, whose  method is based on \cite{cmbfast}. 
As described in section \ref{sec:PertEqs} we have to make
changes in the equations for CDM and baryons and have to include the
modified Klein--Gordon equation for the scalar field. We assume 
a scale--invariant initial power spectrum and adiabatic initial conditions. 

What are the expected effects of the non--vanishing coupling?  There
are two effects, which come in when $\alpha$ differs from zero.
Firstly, as mentioned in section \ref{sec:Cosmo}, the dependence of
the matter density (both for baryons and CDM) on the scale factor is
no longer given by $\rho_{\rm matter} \propto a^{-3}$, but rather by
equation (\ref{matterscaling}). Therefore, relative to the
$\Lambda$CDM model the density of matter (and in particular of
baryons) is {\it larger} at decoupling (we remind the reader that the
models are normalized such that their matter density is equal at the
present epoch). Hence, the amplitudes of the peaks are larger, since a
larger baryon content implies larger amplitudes of the peaks.

Additionally, there are changes to the sound speed $c_s^2 = \dot
p_\gamma/(\dot \rho_{\rm b}+\dot \rho_\gamma)$.  The sound speed
decreases with increasing $\alpha$. This effects the peak separation,
which is proportional to \cite{durrer}
\begin{equation}
\delta l = \frac{\chi (\tau_0 - \tau_{\rm dec})}{D_s},
\end{equation}
where $\chi (\tau_0 - \tau_{\rm dec})$ is the angular diameter
distance to the surface of last scattering and $D_s = \int^{\tau_{\rm
dec}} c_s d\tau $ is the sound horizon.  Aside from having an effect on
$c_s$, the presence of the scalar field does also affect the distance
to the last scattering surface.  Let us estimate the position of the
first peak, given by $l_0 = \pi \chi(\tau_0 - \tau_{\rm dec})/D_s$.
It is simple to calculate $\chi (\tau_0 - \tau_{\rm dec})$:
\begin{equation}
\chi (\tau_0 - \tau_{\rm dec}) = \int_{\tau_{\rm dec}}^{\tau_0} d\tau
\approx \frac{2}{H_0} \frac{1}{1+C}
\end{equation}
where $C=2\alpha^4/(1-8\alpha^2)(1+2\alpha^2)$.
Putting $c_s^2 \approx 1/3$ and neglecting the radiation dominated 
epoch, one finds 
\begin{equation}
D_s \approx \frac{1}{3} \frac{1}{H_0} \frac{2}{1+C} 
a_{\rm dec}^{\frac{1}{2} + \frac{C}{2}}.
\end{equation}
Therefore, the position of the peak is given by (note that 
we have $a(\tau_0) = 1$) 
\begin{equation}\label{elcrude}
l_0 = {\tilde l}_0 a_{\rm dec}^{- C/2},
\end{equation}
where ${\tilde l}_0 \approx 200$ is the position of the peak for the
flat model with no coupling.  Therefore, the peak moves towards larger
values (i.e.\ to the right in the spectrum).

To reach this conclusion we have assumed that $c_s \approx 1/3$, which
is a crude assumption.  Note, however, that if one takes the baryons
into account the $c_s$ is smaller and hence, the sound horizon $D_s$
is smaller.  Thus, $l_0$ is if anything somewhat larger than the value
given in equation (\ref{elcrude}).

\begin{figure}[!ht]
\hspace{1.40cm}\psfig{file=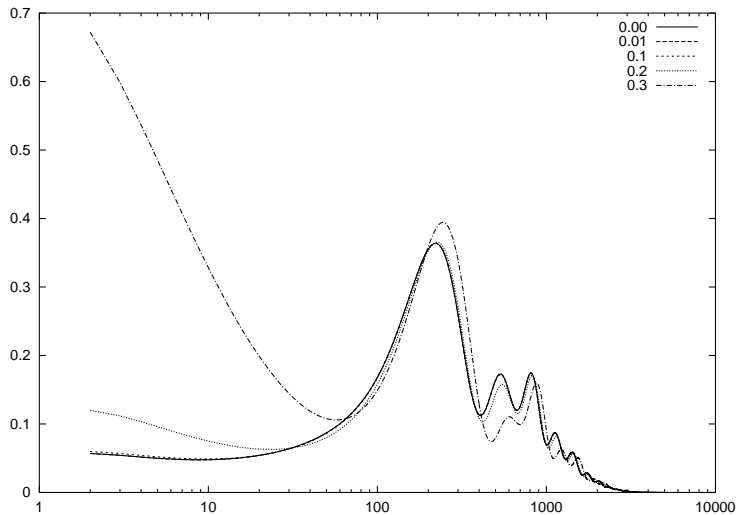,width=10.0cm,angle=270}\vspace{0.75cm}
\caption[h]{
  The temperature anisotropy power spectrum, $l(l+1)C_l/2\pi$, for the
  constant coupling case: the values in the legend are the values of
  $\alpha$.}
\label{fig:constTT}
\end{figure}

The second important change when the coupling is non--vanishing is
that there is a larger contribution from the integrated Sachs--Wolfe
effect.  Even in the matter dominated era the gravitational potential
changes with time, because the matter density contrast no longer grows
like $\delta(\tau) \propto a(\tau)$, as discussed in the last section.
As  the time--evolution of the scale factor is altered in the
presence of the scalar field, the duration of the recombination era is
affected, too.

In figure \ref{fig:constTT} we plot the results for the temperature
anisotropy power spectra.  We assume an initial power spectrum with
equal power on all scales. It can clearly be seen that the both the
amplitude and the separation of the peaks is affected with increasing
coupling. Also, on large scales (low $l$) one can see the additional
power due to the integrated Sachs--Wolfe effect.

\begin{figure}
\hspace{1.40cm}\psfig{file=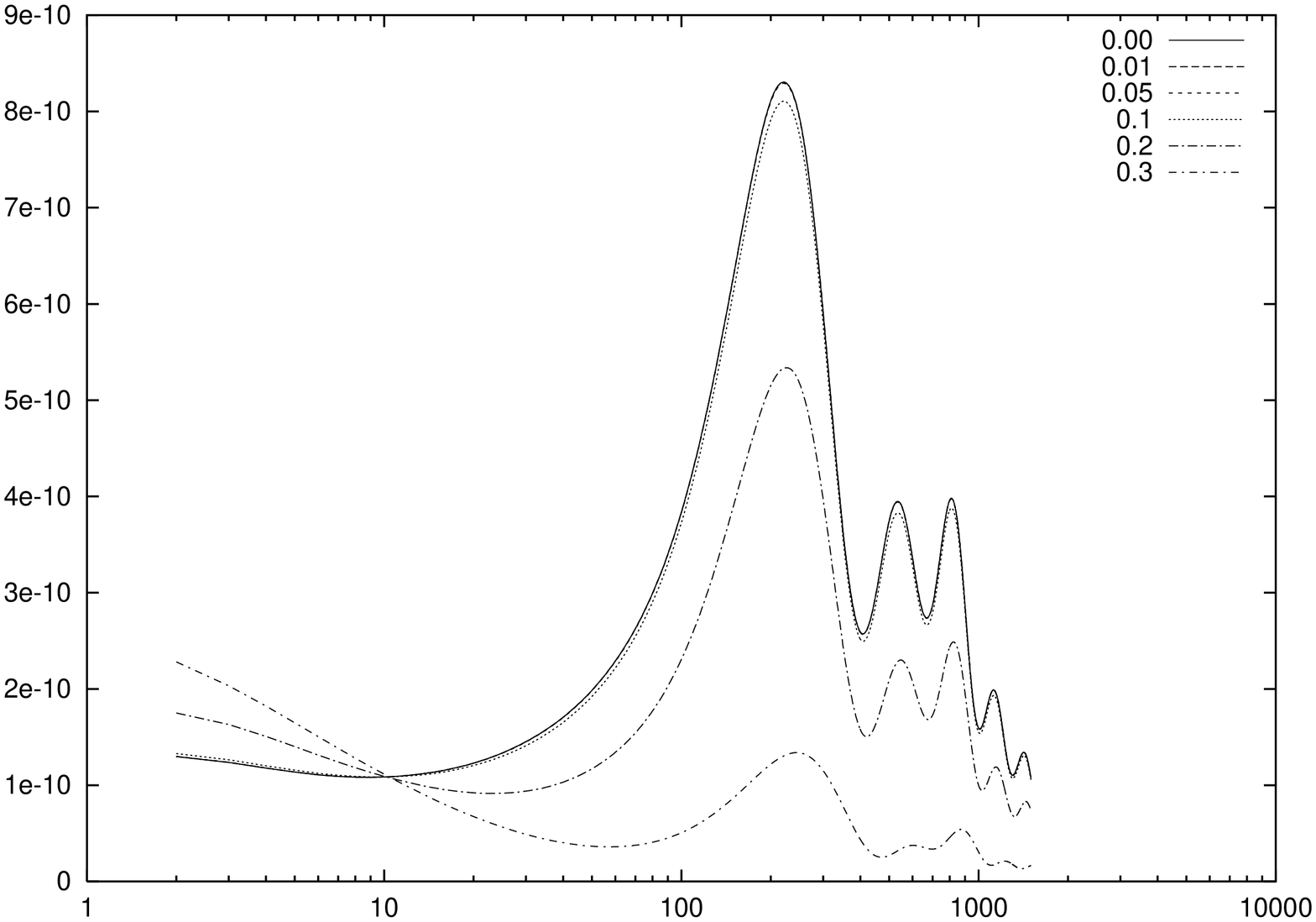,width=9cm,angle=270}\psfig{file=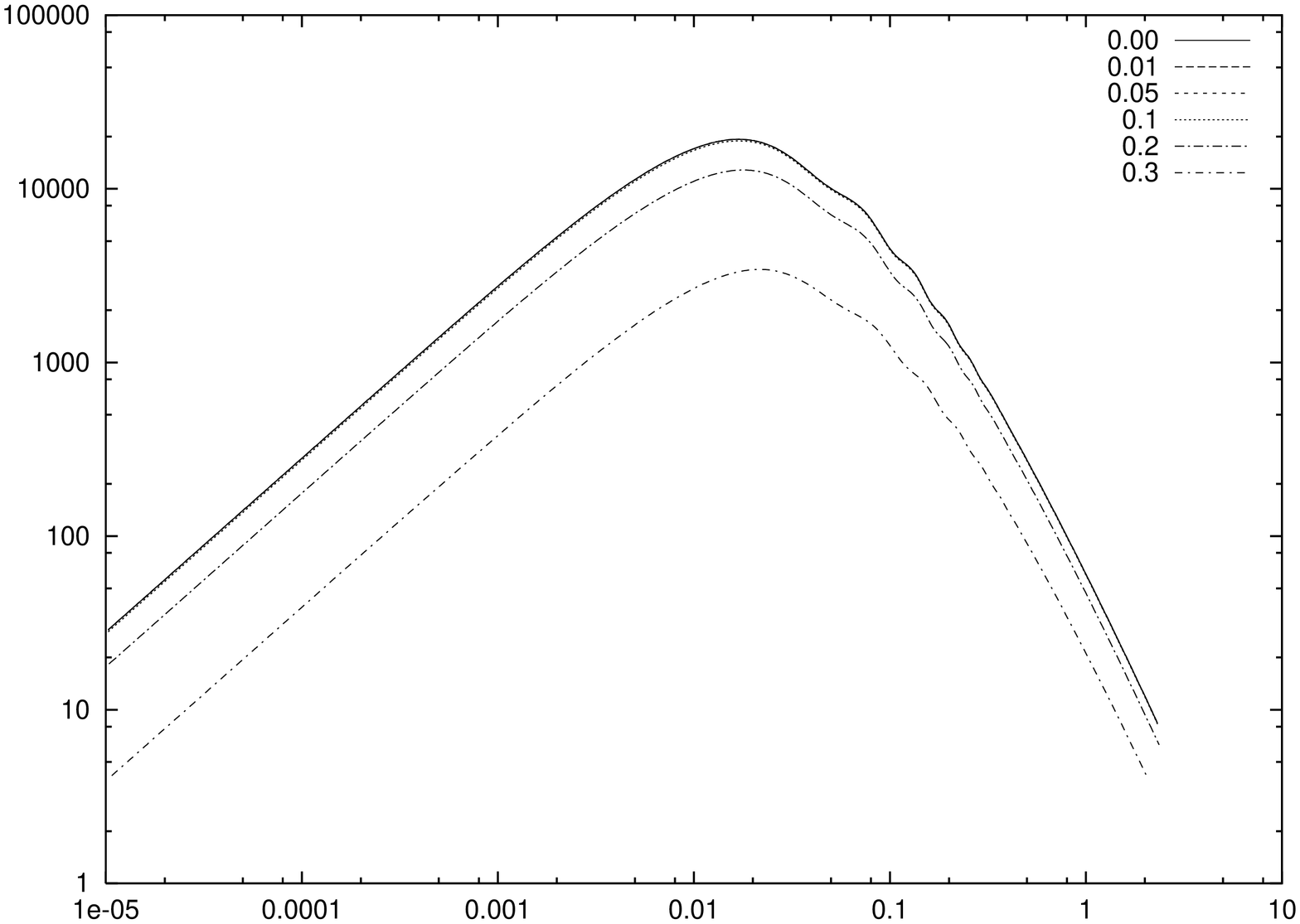,width=9cm,angle=270}\vspace{0.75cm}
\caption[h]{
  COBE-normalized temperature anisotropy $l(l+1)C_l/2\pi$ (left panel)
  and matter power spectrum (right panel) for the case of
  constant coupling. On small scales the COBE--normalized spectra are
  below the predictions for vanishing coupling due to the enhanced
  ISW.}
\label{fig:constCOBE}
\end{figure}

In order to compare the theory to observations, we normalize the
curves appropriately.  Usually this done at large angular scales,
using the COBE data as reference.  The curves normalized to COBE are
plotted in figure \ref{fig:constCOBE}.  Clearly, the normalized curves
show {\it less} power on small scales because of the enhanced
integrated Sachs--Wolfe effect in models with non--vanishing coupling.
On larger scales (small multipole number $l$) one clearly sees a boost
of power. Additionally, the normalized matter power spectra for models 
with non--vanishing coupling are below the $\Lambda$CDM curve.

\begin{figure}[!ht]
\hspace{1.40cm}\psfig{file=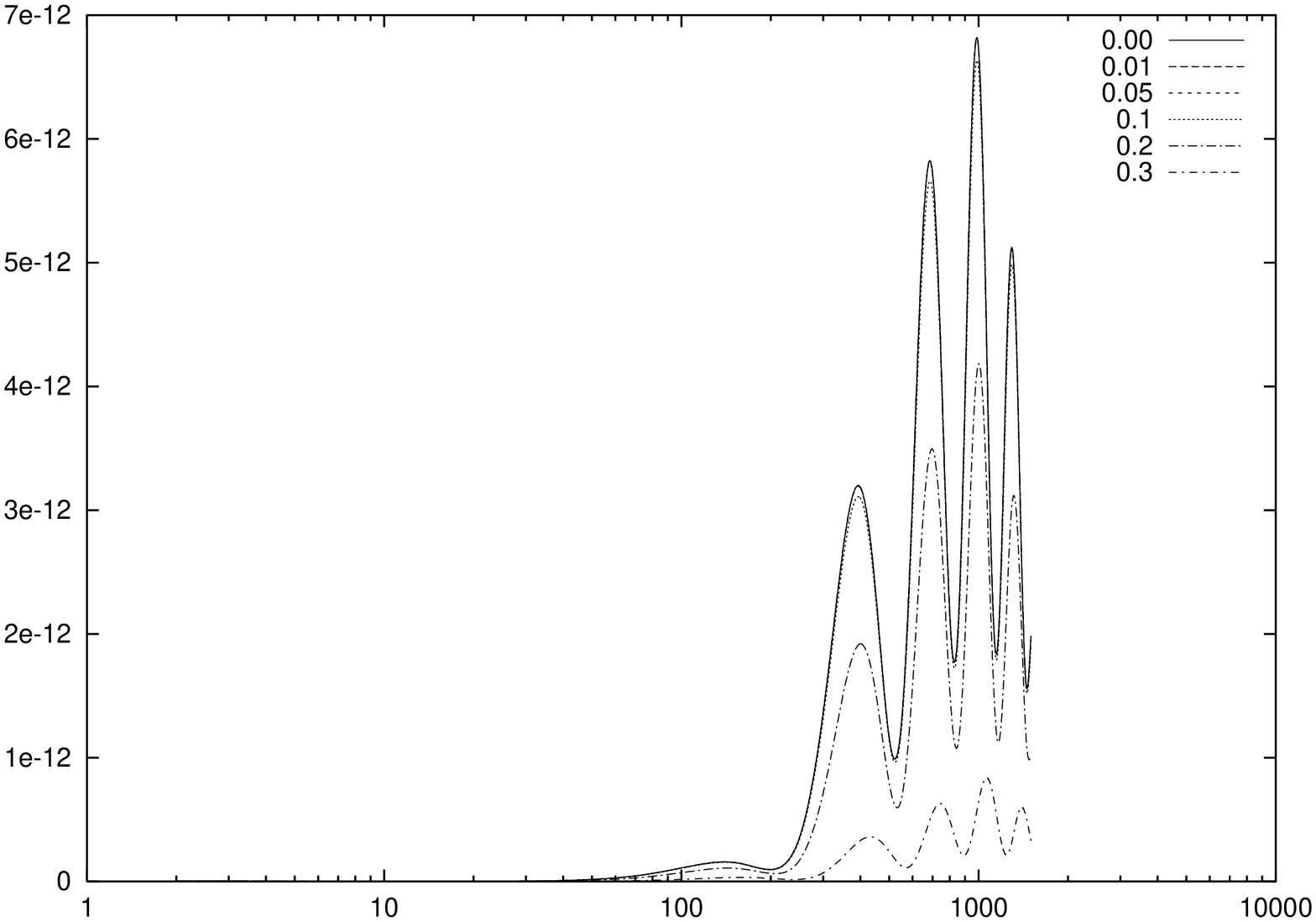,width=9cm,angle=270}\psfig{file=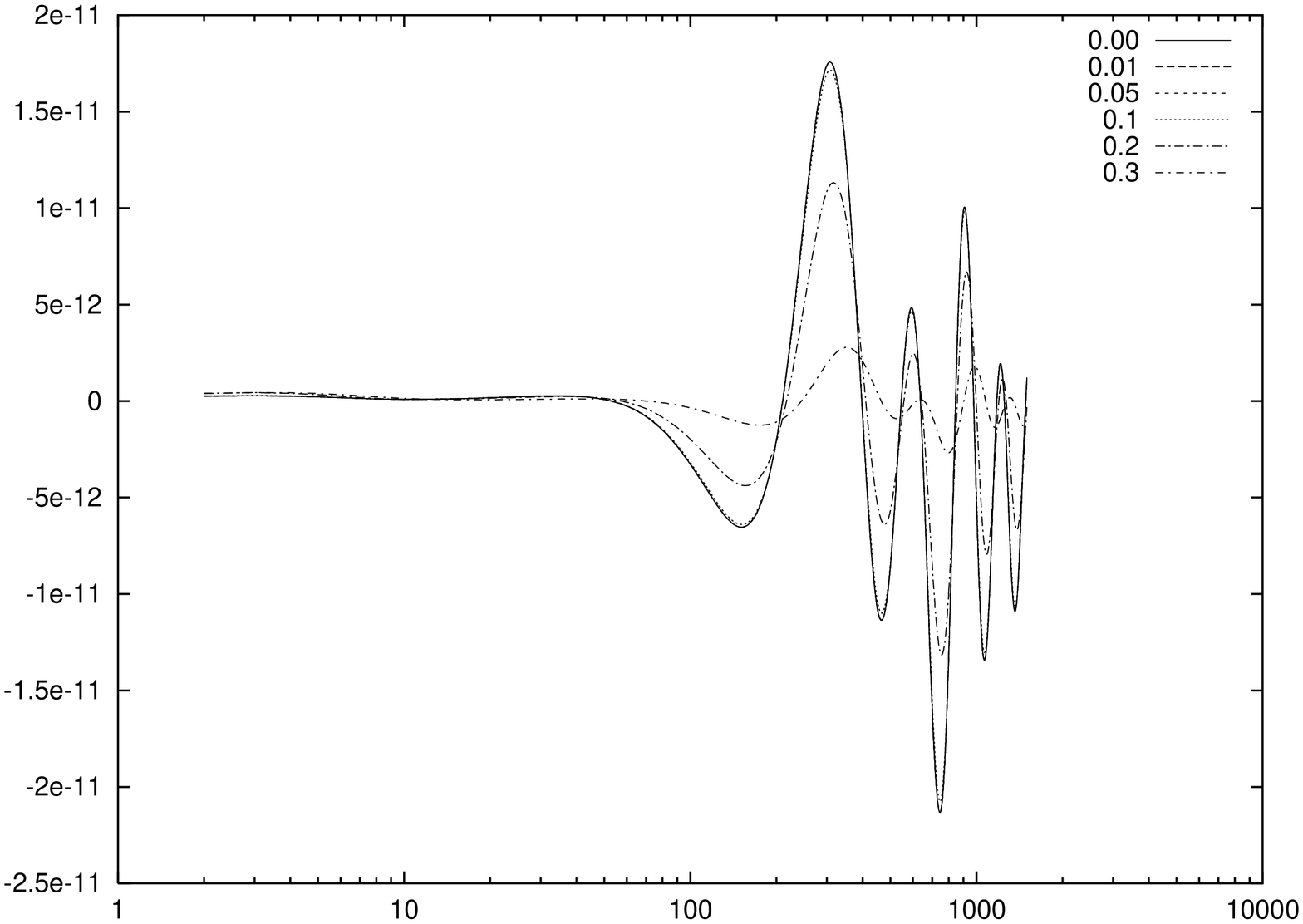,width=9cm,angle=270}\vspace{0.75cm}
\caption[h]{
  COBE-normalized E-mode polarization anisotropy (left panel) and TE
  cross-correlation (right panel) for the case of constant coupling.}
\label{fig:polarizationphi}
\end{figure}

For completeness we plot the E--type polarization spectra and the 
cross correlation between temperature and E--type polarization in 
figure \ref{fig:polarizationphi}. One can clearly see that in 
both spectra on small scales the power is suppressed when $\alpha$ 
is increased.  

\begin{figure}[!ht]
\hspace{1.40cm}\psfig{file=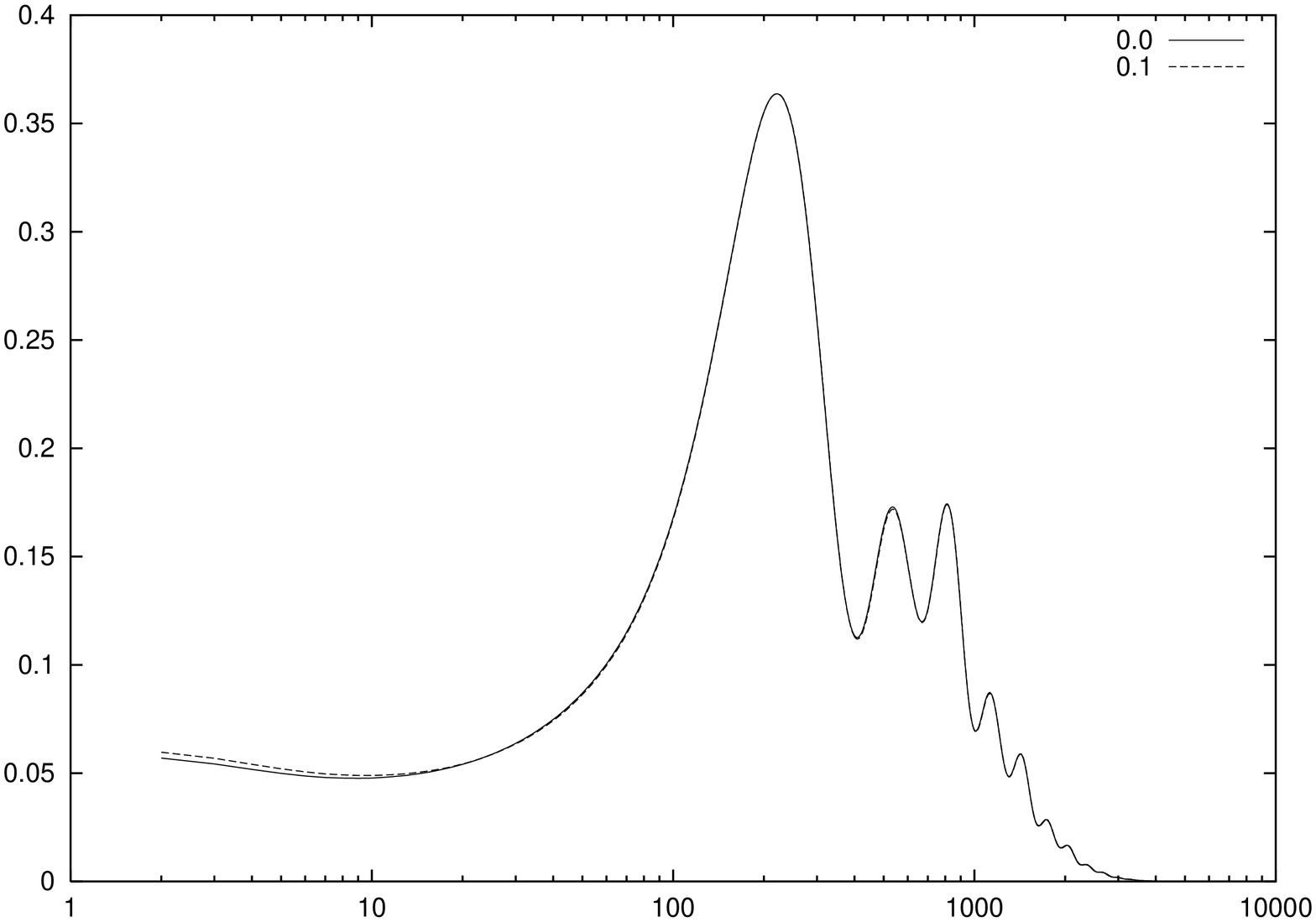,width=9cm,angle=270}\psfig{file=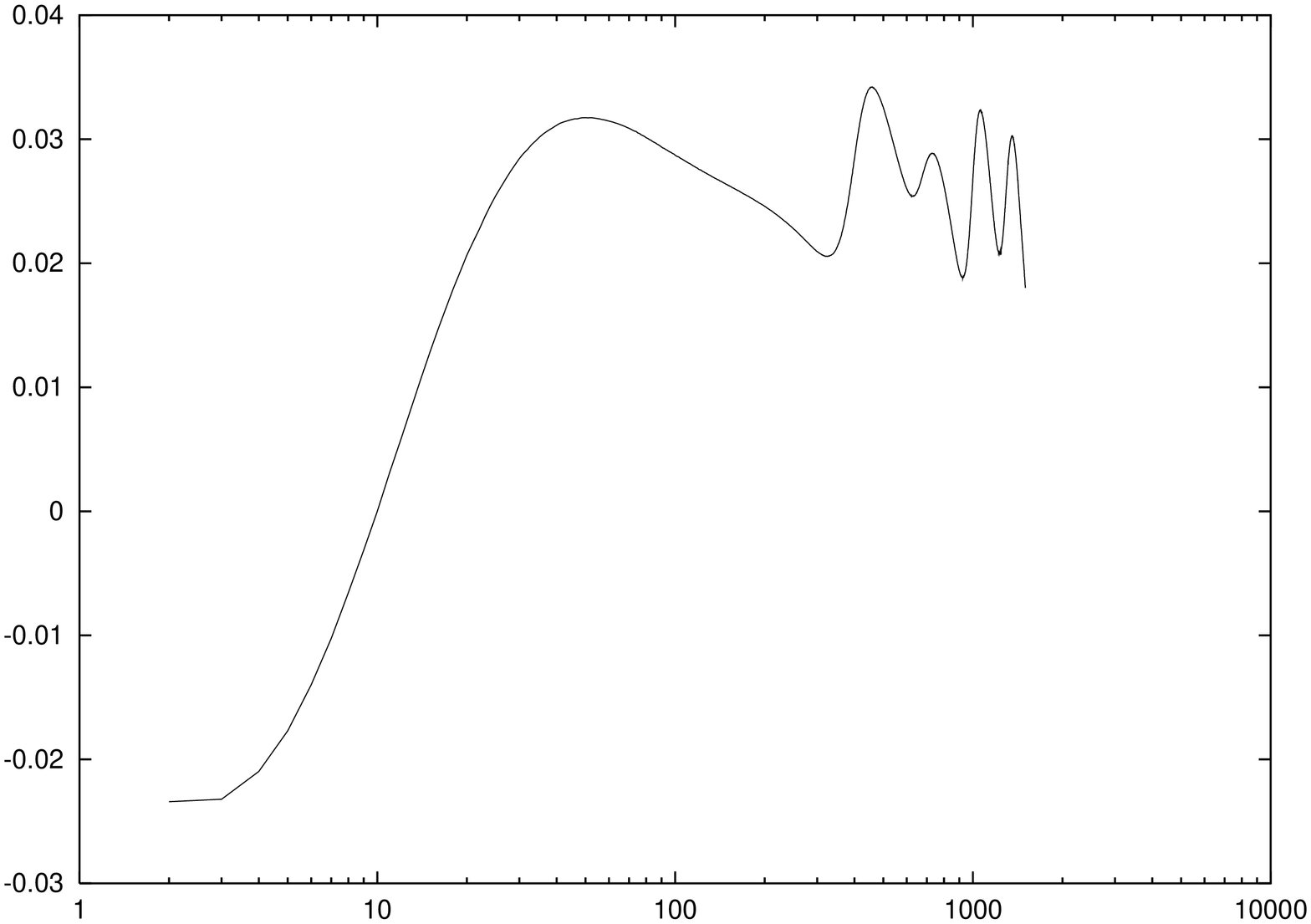,width=9cm,angle=270}\vspace{0.75cm}
\caption[h]{
  CMB anisotropy spectra (left panel) for the constant-coupling case,
  with $\alpha$ taking the extreme value permitted by nucleosynthesis,
  with a reference curve from the $\Lambda$CDM model; the right panel
  shows the relative discrepancy. Although the differences are around 
  3 percent at high multipoles, a different choice of normalization --
  normalizing to COBE at $l=800$, say -- would move all the observed
  discrepancies into the low--multipole region, where cosmic variance 
  dominates.}
\label{fig:bbnconstphi}
\end{figure}

In figure \ref{fig:bbnconstphi} we plot the anisotropy spectra for the $\Lambda$CDM 
model and the model with the maximal value of $\alpha$ allowed by nucleosynthesis 
($\alpha = 0.1$). One can see that the COBE normalized curves differ from each other 
by a few percent on small angular scales.

\subsubsection{Non--constant Coupling} 
In this case, where the field $\phi$ corresponds to the $R$ field in
equation \ref{fundamental}, the coupling function now decays. But also
here, the baryon density is larger at decoupling compared to the
$\Lambda$CDM, which leaves a similar effect on the peak position and
amplitudes as in the case of constant coupling.

\begin{figure}[!ht]
\hspace{1.40cm}\psfig{file=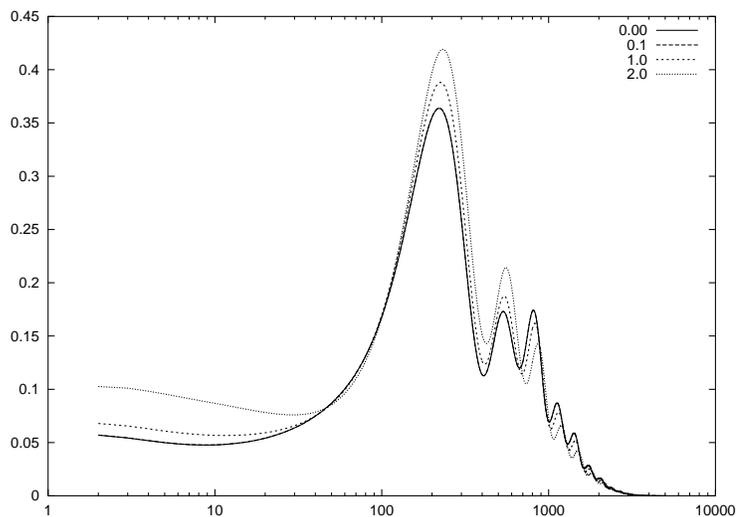,width=10.0cm,angle=270}\vspace{0.75cm}
\caption[h]{
  The temperature anisotropy power spectrum, $l(l+1)C_l/2\pi$, for the
  field-dependent coupling case: the values in the legend are for the
  initial values of the scalar field $\phi$.}
\label{fig:Runnormalized}
\end{figure}

However, well inside the matter dominated epoch, the coupling becomes
very small, so that the density contrast grows essentially like
$a(\tau)$ from that time on. Therefore, the ISW effect is not as
pronounced as in the case of constant coupling. Only if $\phi$ was
initially very large will the field have not evolved to small values
by today; consequently, for the field-dependent coupling case, only
for large initial field values will the effect on the ISW become more
apparent (see figure \ref{fig:Runnormalized}).

\begin{figure}[!ht]
\hspace{1.40cm}\psfig{file=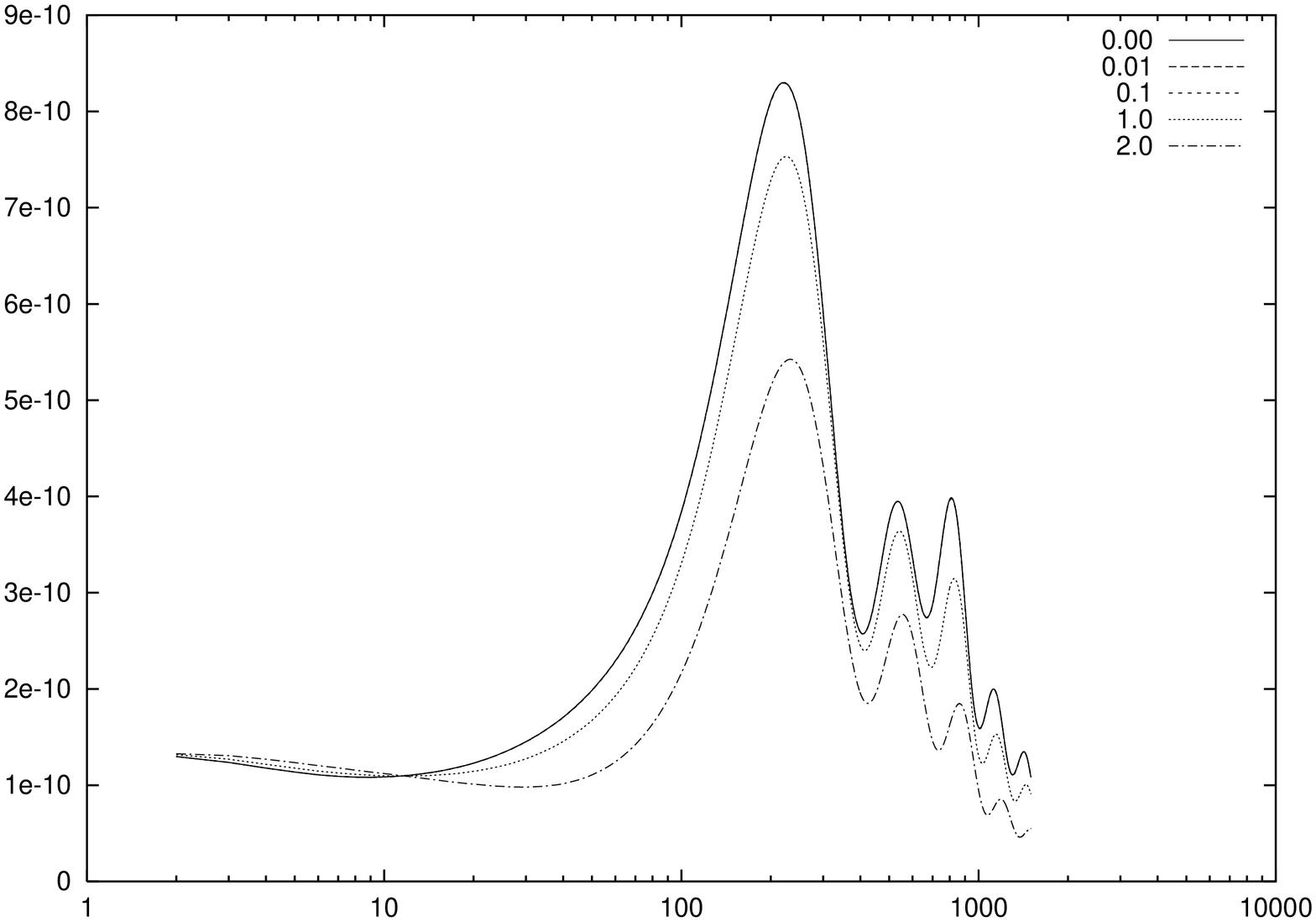,width=9cm,angle=270}\psfig{file=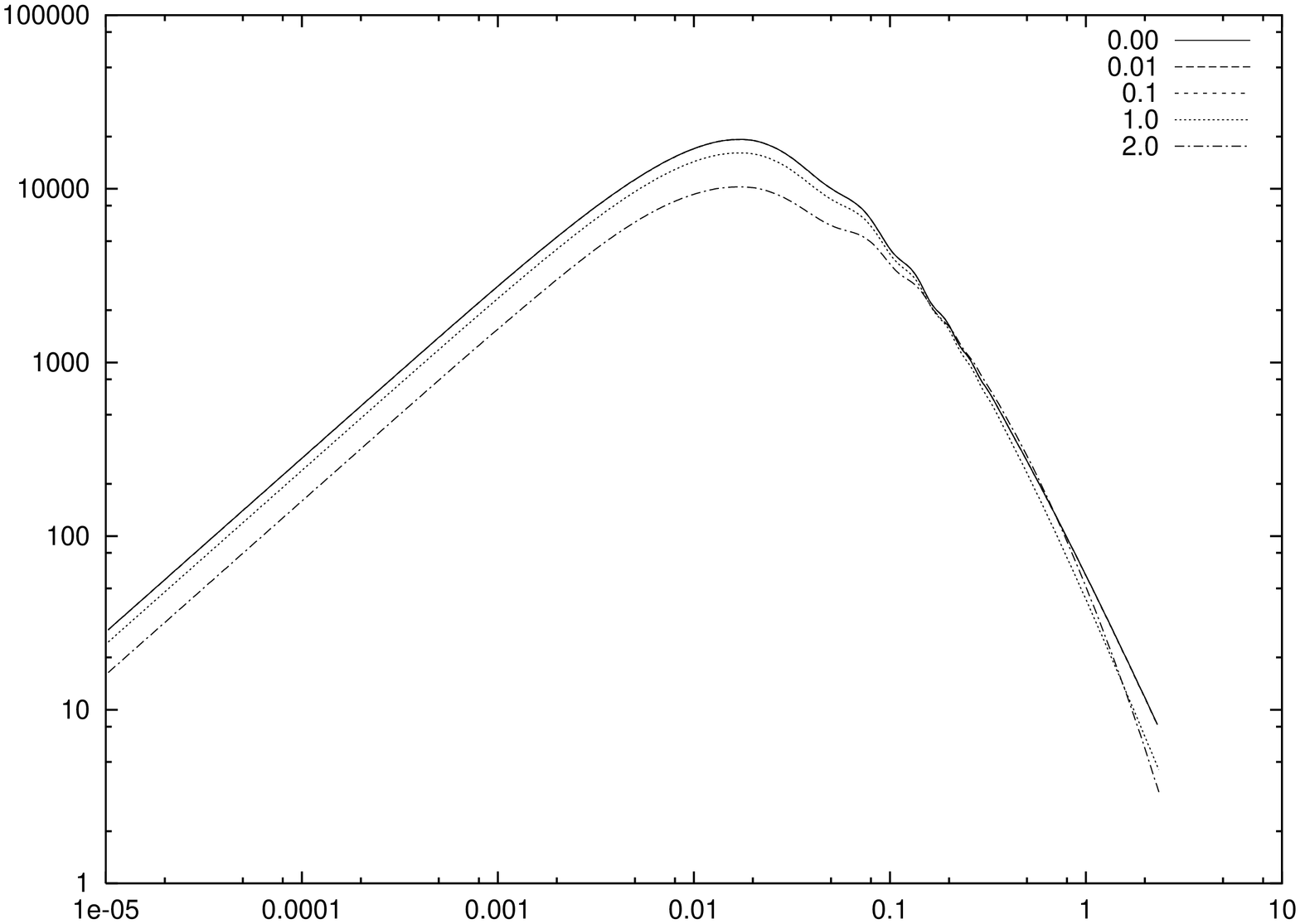,width=9cm,angle=270}\vspace{0.75cm}
\caption[h]{
  COBE-normalized temperature anisotropy $l(l+1)C_l/2\pi$ (left panel)
  and matter power spectrum (right panel) for the case of
  field-dependent coupling.  Similar to the case in figure
  \ref{fig:constCOBE}, on small scales the COBE--normalized spectra
  are below the predictions for vanishing coupling.}
\label{fig:Rnormalized}
\end{figure}

\begin{figure}[!ht]
\hspace{1.40cm}\psfig{file=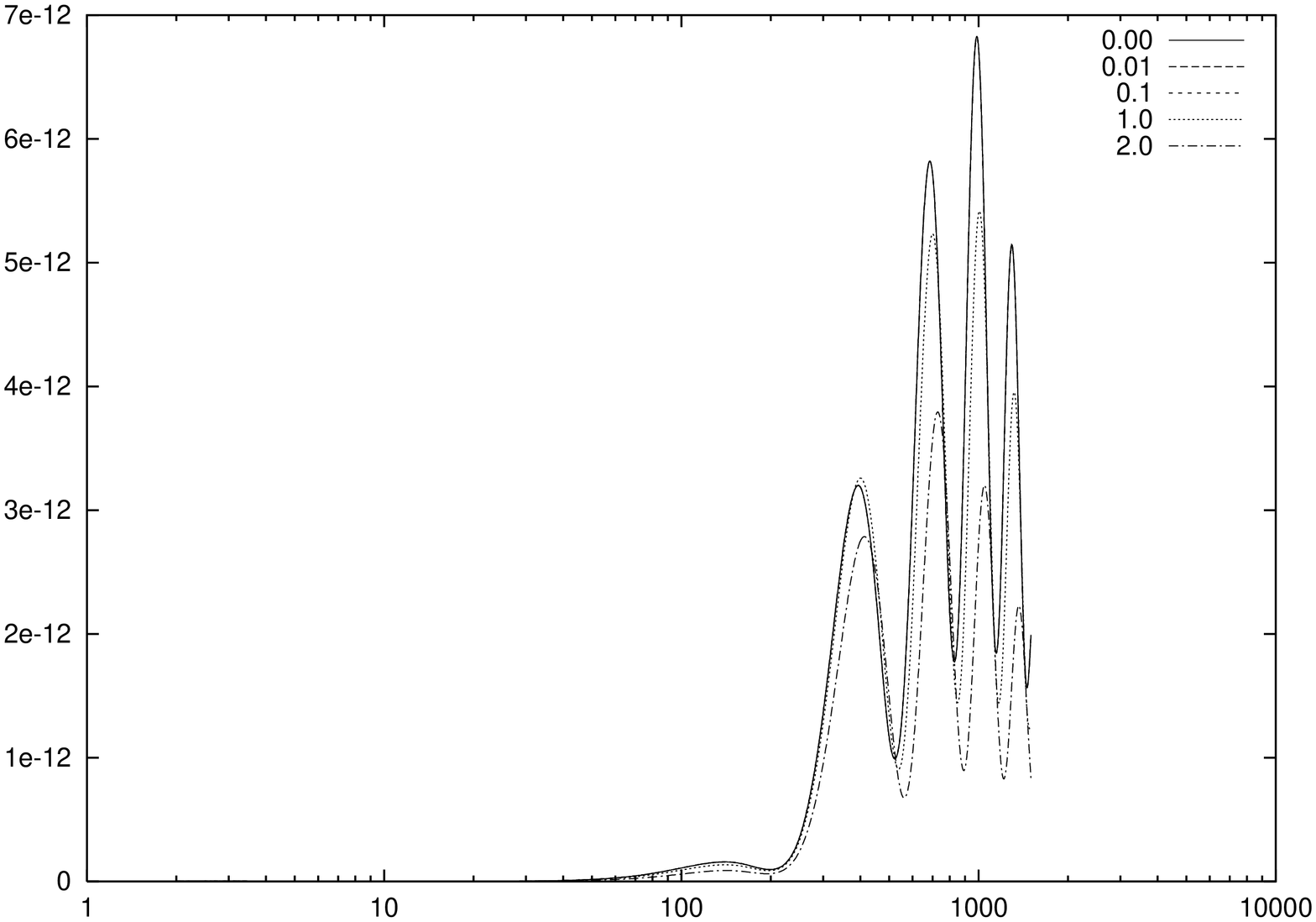,width=9cm,angle=270}\psfig{file=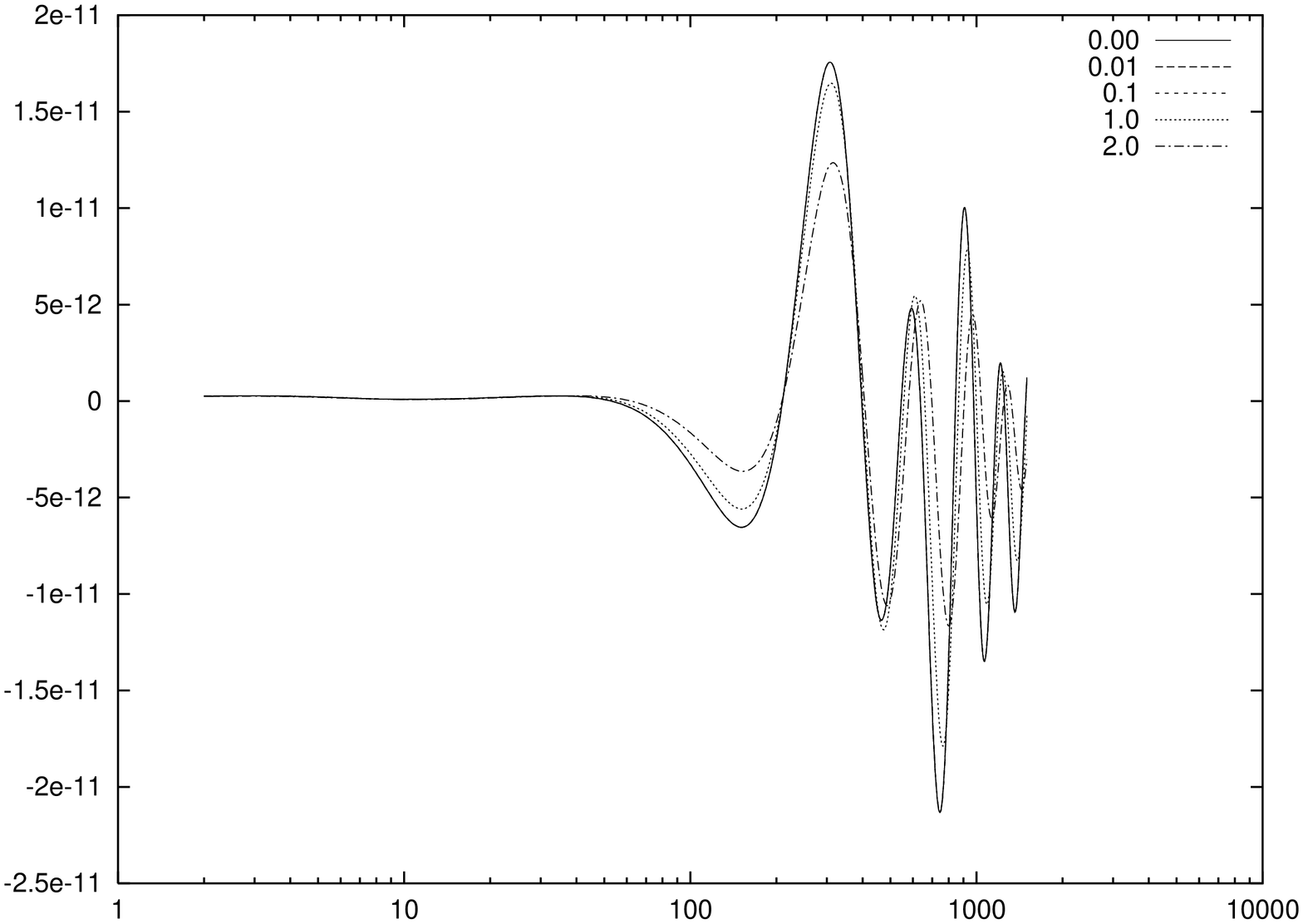,width=9cm,angle=270}\vspace{0.75cm}
\caption[h]{
  COBE-normalized E-mode polarization anisotropy (left panel) and TE
  cross-correlation (right panel) for the case of field-dependent
  coupling.}
\label{fig:Rpol}
\end{figure}
The normalized anisotropy and matter power spectra are shown in figure 
\ref{fig:Rnormalized}, whereas the normalized E-mode polarization 
anisotropy and the TE--cross--correlation are shown in figure \ref{fig:Rpol}.
Increasing the initial value of the field implies that the power 
in both the E--polarization spectrum and the cross correlation spectrum 
between temperature and E--polarization is suppressed on small scales. 

\begin{figure}[!ht]
\hspace{1.40cm}\psfig{file=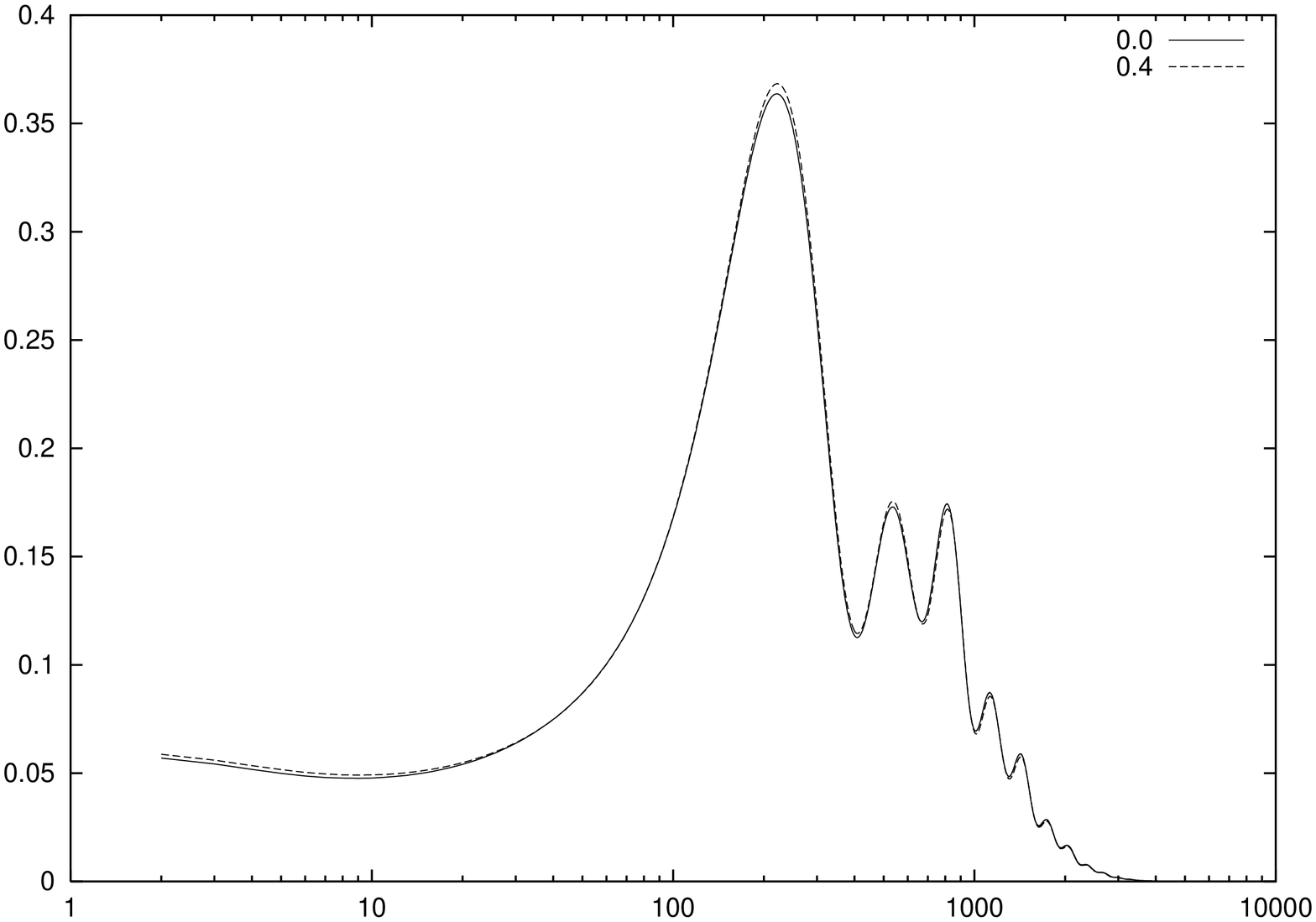,width=9cm,angle=270}\psfig{file=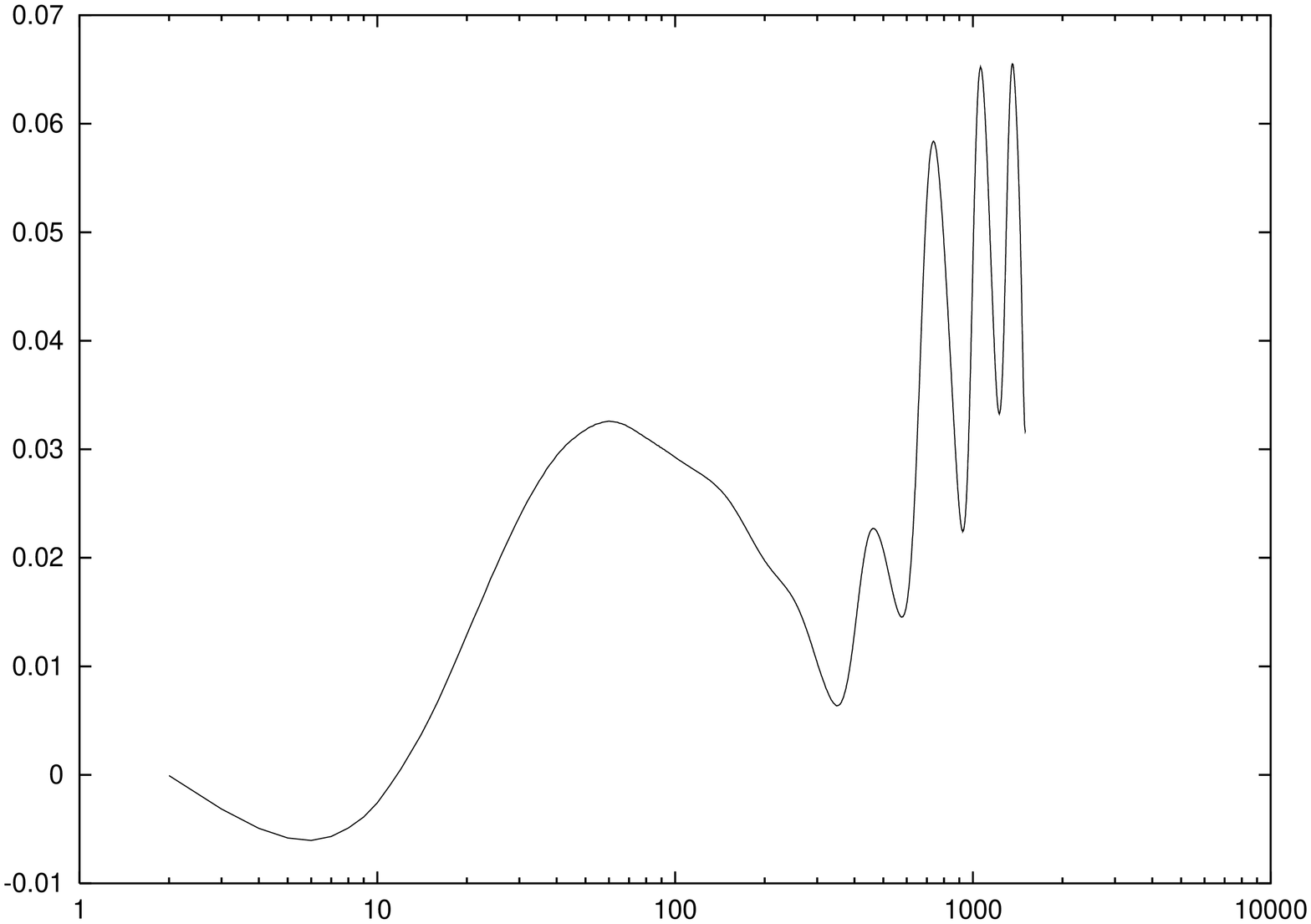,width=9cm,angle=270}\vspace{0.75cm}
\caption[h]{
  CMB anisotropy spectra (left panel) for the case of an initial value
  $(\phi_{\rm ini} = 0.4)$ allowed by nucleosynthesis. The
  right panel shows the relative error between the $\Lambda$CDM and
  the model with $\phi_{\rm ini} = 0.4 $.  For $l\approx
  100$ the difference between the model is as large as 3 percent and
  gets larger at higher $l$ (smaller angular scales).}
\label{fig:bbnconstR}
\end{figure}

The initial field values in figures (\ref{fig:Runnormalized}) and
(\ref{fig:Rnormalized}) have been used without taking into account
constraints from nucleosynthesis (\ref{nucleosynthesis}), which
implies that $\phi_{\rm nuc}<0.4$.  In figure
\ref{fig:bbnconstR} we plot the cases for $\phi_{\rm init} =
0.4$ and the $\Lambda$CDM curve.  One can see that there are
differences in the predictions for the temperature anisotropy as large
as 6 percent at small angular scales and 3 percent on degree
scale. This is potentially distinguishable with WMAP and the Planck
Surveyor.

We would like to point out that constraints on $\phi$ become stronger,
once the CMB is included. In all cases plotted in the figures, the
field fulfills the constraints today. However, a large field value in
the early universe modifies the predictions for the temperature and
polarization power spectra (as well as the matter power spectra).
Hence, the CMB provides vital {\it complementary} information to the
local experiments.

\section{Conclusion}
\label{sec:Conc}

In this paper we have investigated the implications of time--varying
extra dimensions for CMB anisotropies and large scale structure
formation. In theories with extra dimensions, certain scalar fields,
so called `moduli fields', appear in the low energy effective theory
coupling directly to matter fields.  We have investigated in
particular the couplings appearing in a five--dimensional model based
on the brane world idea. The model contains two branes which are the
boundaries of the higher--dimensional spacetime, with, in general,
four-dimensional matter fields confined on these branes.
Additionally, we allowed for a scalar field propagating in the
extra--dimensional spacetime.  At low energies the dynamics of this
system is described by a bi--scalar--tensor theory.  One of the scalar
fields has a constant (i.e.\ field-- and coordinate--independent)
coupling to the matter fields, whereas in the case of the second field
the coupling function is field dependent.

We have investigated the effects of the individual fields on the CMB
anisotropies, the cross--correlation between temperature and
polarization of the CMB and the matter power spectrum. Due to the
coupling, both the amplitude and the positions of the peaks are
affected. The spectra normalized to COBE lie below the spectra of the
reference $\Lambda$CDM model on small scales.

The results presented in section \ref{sec:Pert} imply that, for the
brane world model presented in section \ref{sec:Effective},
potentially there is a measurable effect of both fields on CMB. If it
turns out that the second field needs to be already quite small during
the radiation dominated epoch, then some mechanism has to drive it to
these small values in the very early universe. This can happen, for
example, if there is some matter on the second brane. In this case,
the field $R$ is driven even faster to zero, so that it might be small
at decoupling.  Note, however, that in this case the evolution of
perturbations are affected in the radiation dominated epoch.

Clearly, the CMB anisotropies as well as the polarization provide {\it
complementary} information about time--evolution of the extra
dimensions.  Together with other tests, such as gravity experiments in
the solar system and/or on earth and tests of time--varying gauge
couplings and primordial nucleosynthesis, these cosmological
considerations strongly constrain any model involving extra
dimensions. For example in the case of the model considered here, the
field with constant coupling is already constrained by local
experiments such that its effects on the CMB are negligible. For the
second field, however, we have seen that although its effect is
constrained to be minimal today, it can have a big impact on the CMB
anisotropies.

There is another important point not addressed in this paper: we have
not discussed the time evolution of Kaluza--Klein excitations and
their influence on the CMB.  They might still be present in the
radiation dominated epoch, thereby affecting the evolution of
perturbations at least initially.  Their presence could imply, for
example, that the initial perturbations are not purely adiabatic.  The
details would depend on the mechanism which produces the primordial
fluctuations, such as inflation or the cyclic model, see
e.g.\ \cite{neil1} and \cite{neil2}. To address this important issue
one has to go beyond the moduli space approximation (and in fact
beyond the lowest order in the method used in \cite{kannosoda},
\cite{kannosoda2} and \cite{shiromizu}).

In this paper we have investigated the effects of moduli fields on the
CMB motivated from a certain class of brane world theories. There are
other brane world models, in which the theory at low energies cannot
be described by an action of the form (\ref{fundamental}). For example,
in the models of \cite{Gregory}--\cite{Dvali2}, gravity becomes
five--dimensional at large distances. We expect that some of the
conclusions drawn in this paper do not hold for such
models. Nevertheless, the CMB gives useful constraints on these
models, too \cite{Silk}.

\vspace{1cm}

{\bf Acknowledgements:} We are grateful to Constantinos Skordis and
Domenico Tocchini--Valentini for useful discussions and comments and
to Anthony Lewis for providing his CAMB quintessence code.  We
acknowledge support from the Anglo--French alliance exchange
programme.  This work is supported in part by PPARC.  P.B.\ is
partially fundedd by the RTN European programme HPRN-CT-2000-00148.


\begin{thebibliography}{99}
\bibitem{liddlelyth} A.R. Liddle and D.H. Lyth, {\it Cosmological Inflation 
and Large Scale Structure}, Cambridge University Press (2000)
\bibitem{peacock} J. Peacock, {\it Cosmological Physics}, 
Cambridge University Press (1999)
\bibitem{dodelson} S. Dodelson, {\it Modern Cosmology}, Academic Press (2003)
\bibitem{WMAP} C.L. Bennett, M. Halpern, G. Hinshaw, N. Jarosik, A. Kogut, 
M. Limon, S.S. Meyer, L. Page, D.N. Spergel, G.S. Tucker, E. Wollack, E.L. Wright, 
C. Barnes, M.R. Greason, R.S. Hill, E. Komatsu, M.R. Nolta, N. Odegard, H. Peiris, 
L. Verde and J.L. Weiland, astro-ph/0302207
\bibitem{fujiimaeda} Y. Fujii and K-I Maeda, 
{\it The Scalar--Tensor Theory of Gravitation}, Cambridge University Press (2003)
\bibitem{kamionkowski} X-L Chen and M. Kamionkowski, Phys. Rev. D {\bf 60}, 104036 (1999)
\bibitem{uzan} A. Riazuelo and J-P Uzan, Phys. Rev. D {\bf 66}, 023525 (2002)
\bibitem{chiba} R. Nagata, T. Chiba and N. Sugiyama, Phys. Rev. D {\bf 66}, 103510 (2002)
\bibitem{amendola} L. Amendola, C. Quercellini, D. Tocchini-Valentini and A. Pasqui, 
ApJL {\bf 583}, L53 (2003)
\bibitem{bean} R. Bean, Phys. Rev. D{\bf 64}, 123516 (2001) 
\bibitem{zahn} O. Zahn and M. Zaldarriaga, Phys. Rev. D {\bf 67}, 063002 (2003)
\bibitem{polchinski} J. Polchinski, {\it String Theory}, Cambridge University Press (1999)
\bibitem{braxvandebruck} Ph. Brax and C. van de Bruck, Class. Quant. Grav. {\bf 20}, 
R201 (2003)
\bibitem{langlois} D. Langlois, Prog. Theor. Phys. Suppl.{\bf 148}, 181 (2003)
\bibitem{maartens} R. Maartens, Prog. Theor. Phys. Suppl.{\bf 148}, 213 (2003) 
\bibitem{RSI} L. Randall and R. Sundrum, Phys. Rev. Lett.{\bf 83}, 3370 (1999)
\bibitem{damour} T. Damour and K. Nordtvedt, Phys. Rev. D {\bf 48}, 3436 (1993)
\bibitem{lukas} A. Lukas, B.A. Ovrut, K.S. Stelle and D. Waldram, Nucl. Phys. B {\bf 552}, 
246 (1999)
\bibitem{binetruy} P. Binetruy, C. Deffayet and D. Langlois, 
Nucl. Phys. B {\bf 565}, 269 (2000) 
\bibitem{boundaryinflation} A. Lukas, B.A. Ovrut and D. Waldram, 
Phys. Rev. D {\bf 61}, 023506 (2000)
\bibitem{clinevinet} J.M. Cline and J. Vinet, JHEP {\bf 0202}, 042 (2002) 
\bibitem{modulipaper} Ph. Brax, C. van de Bruck, A.-C. Davis and C.S. Rhodes, 
Phys. Rev. D {\bf 67}, 023512 (2003)
\bibitem{braxdavis} Ph. Brax and A.-C. Davis, Phys. Lett. B {\bf 497}, 289 (2001)
\bibitem{kannosoda} S. Kanno and J. Soda, Phys. Rev. D {\bf 66} 083506 (2002) 
\bibitem{kannosoda2} S. Kanno and J. Soda, hep-th/0303203
\bibitem{kobayashi} S. Kobayashi and K. Koyama, JHEP 0212:056 (2002)
\bibitem{shiromizu} T. Shiromizu, K. Koyama and K. Takahashi, 
Phys. Rev. D {\bf 67}, 084022 (2003)
\bibitem{koyama} K. Koyama, astro-ph/0303108 
\bibitem{RSII} L. Randall and R. Sundrum, Phys. Rev. Lett. {\bf 83}, 4690 (1999)
\bibitem{conformal} V. Faraoni, E. Gunzig and P. Nardone, Fund. Cosmic Phys. 
{\bf 20}, 121 (1999)
\bibitem{Will} C. Will, Living Rev.Rel. {\bf 4} 4 (2001) 
\bibitem{bartolo} N. Bartolo and M. Pietroni, Phys. Rev. D {\bf 61} 023518 (1999)
\bibitem{amendola2} L. Amendola, Phys. Rev. D {\bf 62} 043511 (2000)
\bibitem{mabertschinger} C-P Ma and E. Bertschinger, ApJ {\bf 455}, 7 (1995)
\bibitem{durrer} R. Durrer, J. Phys. Stud. {\bf 5}, 177 (2001) 
\bibitem{hudodelson} W. Hu and S. Dodelson, Ann. Rev. Astron. Astrophys. {\bf 40}, 
171 (2002)
\bibitem{CMBshortbegin} R. Durrer, astro-ph/0109274 
\bibitem{zaldareview} M. Zaldarriaga, astro-ph/0305272 
\bibitem{Hushort} W. Hu, Annals Phys. {\bf 303} 203 (2003) 
\bibitem{CMBshortend} E. Wright, astro-ph/0305591
\bibitem{husugiyama} W. Hu and N. Sugiyama, ApJ {\bf 444}, 489 (1995)
\bibitem{lewis} A. Lewis, A. Challinor and A. Lasenby,  ApJ {\bf 538}, 473 (2000)
\bibitem{cmbfast} U. Seljak and M. Zaldarriaga, ApJ {\bf 469}, 437 (1996)
\bibitem{neil1} P.J. Steinhardt and N. Turok, Phys. Rev. D {\bf 65} 126003 (2002) 
\bibitem{neil2} S. Gratton, J. Khoury, P.J. Steinhardt and N. Turok, astro-ph/0301395 
\bibitem{Gregory} R. Gregory, V.A. Rubakov and S.M. Sibiryakov, 
Phys. Rev. Lett. {\bf 84} 5928 (2000)
\bibitem{Dvali} G.R. Dvali, G. Gabadadze and M. Porrati, Phys. Lett. B {\bf 485} 208 (2000)
\bibitem{Dvali2} G.R. Dvali, G. Gabadadze and M. Porrati, Phys. Lett. B {\bf 484} 112 (2000)
\bibitem{Silk} P. Binetruy and J. Silk, Phys. Rev. Lett. {\bf 87} 031102 (2001) 
\end{thebibliography}
\end{document}